\documentclass[submission,copyright,creativecommons]{eptcs}
\usepackage{underscore}  
\usepackage[shortlabels]{enumitem} 
\usepackage{macros}
\usepackage{amsfonts}
\usepackage{graphicx}
\usepackage{float}
\usepackage{mathtools}
\usepackage[table,xcdraw]{xcolor}
\usepackage{listings}
\usepackage{amssymb}
\usepackage{subfigure}
\usepackage{amsfonts}
\usepackage{graphicx}
\usepackage{cleveref}
\usepackage{soul}
\newtheorem{definition}{Definition}
\newtheorem{theorem}{Theorem}
\newtheorem{proposition}{Proposition}
\newtheorem{lemma}{Lemma}
\usepackage{xcolor}
\usepackage{textcomp}
\usepackage{amsfonts}
\usepackage{amssymb}
\usepackage[normalem]{ulem}

\providecommand{\event}{EXPRESS/SOS 2022} 

\usepackage{iftex}

\ifpdf
  \usepackage{underscore}         
  \usepackage[T1]{fontenc}        
\else
  \usepackage{breakurl}           
\fi

\title{Token Multiplicity in Reversing Petri Nets Under the Individual Token Interpretation}
\author{Anna Philippou \quad\quad
Kyriaki Psara
\institute{Department of Computer Science,
	University of Cyprus\\
Nicosia, Cyprus}
\email{annap@ucy.ac.cy\quad
kpsara01@ucy.ac.cy }
}
\def\titlerunning{Token Multiplicity in Reversing Petri Nets Under the Individual Token Interpretation}
\def\authorrunning{A. Philippou \& K. Psara  }
\begin{document}
\maketitle

\begin{abstract}
Reversing Petri nets (RPNs) have recently been proposed as a net-based
approach to model causal and out-of-causal order reversibility. They are
based on the notion of individual tokens that can be connected together via bonds. 
In this paper we extend RPNs by allowing multiple tokens of the same
type to exist within a net based on the individual token interpretation of Petri nets.
According to this interpretation, tokens of the same type are distinguished
via their causal path. We develop a causal semantics of the model
and we prove that  the expressive power of RPNs with multiple tokens is 
equivalent to that of RPNs with single tokens by establishing an isomporphism
between the Labelled Transition Systems (LTSs) capturing the reachable parts
of the respective RPN models. 
\end{abstract}

\section{Introduction}

Reversible computation is a form of computing where transitions can be executed
in both the forward and the reverse direction, allowing systems to return
to past states.  It has been 
attracting increasing attention due to its application in a variety of fields
such as 
low-power computing, biological modelling,  quantum computation, robotics, and 
distributed systems. 

In the sequential setting reversibility is generally understood 
as the ability to execute past actions in the exact inverse order in which they
occurred, a process referred
to as \emph{backtracking}.
However, in the concurrent setting matters are less clear. Indeed, various approaches 
have been investigated within a variety of formalisms~\cite{TransactionsRCCS,phillips2007reversing,LaneseMS16,ConRev,RPNscycles,DBLP:journals/lmcs/MelgrattiMU20}.
One of the most well-studied approaches considered suitable for a wide variety
of concurrent systems is that of \emph{causal-consistent reversibility}
advocating that a transition can be undone only if all its effects, if any,
have been undone beforehand~\cite{RCCS}. The study of reversibility also extends 
to \emph{out-of-causal-order reversibility}, a form of reversing where executed
actions can be reversed in an out-of-causal order~\cite{ERK,Conflict,KUHN201818}
most notably featured in biochemical systems. 

In this work, we focus on Reversing Petri
Nets~\cite{RPNscycles} (RPNs), a reversible model inspired by Petri nets 
that allows the modelling
of reversibility as realised by backtracking, causal-order, and out-of-causal-order
reversing.  A key challenge when reversing computations in Petri
nets is handling \emph{backward conflicts}. These conflicts arise
when tokens occur in a certain place due to different causes
making unclear which transitions ought to be reversed.
To handle this ambiguity, RPNs introduce the notion of a \emph{history}
of transitions, which records causal information of
executions. Furthermore, inspired by biochemical systems as well
as other resource-aware applications, the model 
employs named tokens that can be connected together
to form bonds, and are preserved during execution. 

A restriction
in RPNs is that each token is unique and in order to model
a system with multiple items of the same type, it is necessary
to employ a distinct token for each item, at the expense of the net size.
In the current paper we consider an extension of RPNs, which
allows multiple tokens of the same type. The introduction of multiple identical 
tokens creates further challenges involving
backward conflicts and requires to extend the RPN machinery for
extracting the causal dependencies between transitions. 
We note that formalizing causal dependencies
is a well-studied problem in the context of Petri nets, where various
approaches have been proposed to reason about 
causality~\cite{MissingOne,MissingThree,MissingFour}.
In this work we draw inspiration from the so-called individual token 
and collective interpretations of Petri nets~\cite{ConfStruct,individual}.
The collective token philosophy considers all tokens of a certain
type to be identical,  which results in
ambiguities when it comes to causal dependencies. 
In contrast, in an individual token interpretation, tokens are distinguished based
on their causal path. 
This approach leads to more complicated semantics since to achieve
token individuality requires precise correspondence between the token instances 
and their past. However, it enables backward determinism, which is a crucial 
property of reversible systems.

\paragraph{Contribution.}
In this paper we extend RPNs to support multiple tokens of the same
type following the individual token interpretation. As such,
tokens  are associated with their causal history and,
while tokens of the same type are equally eligible to fire a transition
when going forward, when going backwards they are able to reverse only
the transitions they have previously fired. In this context, we
define a causal semantics for the model, based on the intuition that
a causal link exists between two transitions if
a token produced by one was used to fire the other. This leads
to the observation that a transition may reverse in
causal order only if it was the last transition executed by all the tokens
it has involved. 
We note that this approach allows a causal-order reversible
semantics that, unlike the original RPN model, does not require any global
history information. In fact, all information necessary for reversal
is available locally within the history of tokens.
Subsequently, we turn to study the expressiveness of the presented model
in comparison to RPNs with single tokens. To do this we employ
Labelled  Transition Systems (LTSs) capturing the state space of RPN models.
We show that for any RPN with multiple tokens there exists an RPN with
single tokens with an isomorphic LTS, thereby confirming our conjecture 
that RPNs with single tokens are as expressive as RPNs with multiple tokens.

\paragraph{Related Work.}
The first study of reversible computation within Petri nets was proposed  
in~\cite{PetriNets,BoundedPNs}, where the authors investigated the effects of 
adding \emph{reversed} versions of selected transitions 
by reversing the directions of a transition's arcs. 
Unfortunately, this approach 
to reversibility violates causality.
Towards examining causal consistent reversibility the work in~\cite{Unbounded} 
investigates whether it is possible to add a complete set of effect-reverses for a given 
transition without changing the set of reachable markings, showing that this problem is in general undecidable.
In another line of work~\cite{DBLP:journals/lmcs/MelgrattiMU20} propose a causal semantics for P/T nets by identifying the 
causalities and conflicts of a P/T net through unfolding it into an equivalent occurrence 
net and subsequently introducing appropriate reverse transitions to create a coloured Petri
net (CPN) that captures a causal-consistent reversible semantics. 
On a similar note,~\cite{RON} introduces the notion of reversible occurrence nets and 
associates a reversible occurrence net to a causal reversible prime event structure, and 
vice versa. Finally,~\cite{Indi} introduces
a reversible approach to Petri nets following the 
individual token interpretation. This 
work is similar to our approach though it refers to a basic PN model, which does not contain 
named tokens nor bonds, and it does not support backtracking and 
out-of-causal reversibility.

The modelling of bonding in the context of reversibility was first considered
within reversible processes and event structures in~\cite{Bonding}, where its
usefulness was illustrated with examples taken from software engineering and 
biochemistry. Reversible frameworks that feature bonds as first-class
entities, like RPNs, also include the Calculus of 
Covalent Bonding~\cite{KUHN201818}, which supports causal and 
out-of-causal-order reversibility in the 
context of chemical reactions, as well as the Bonding Calculus~\cite{NaCo18},
a calculus developed for modeling covalent bonds between molecules in
biochemical systems. In fact, the latter two frameworks and RPNs were
reviewed and compared for modeling chemical reactions 
in~\cite{RCbook} with case study the autoprotolysis 
of water. 

This paper extends a line of research on reversing Petri nets, initially 
introduced for acyclic nets~\cite{RPNs}  
and subsequently for nets with cycles~\cite{RPNscycles}. The usefulness
of the framework was illustrated in a number of examples including
the modelling of long-running transactions with compensation and a 
signal-passing mechanism used by the ERK pathway. 
The RPN framework has been extended to control 
reversibility in~\cite{RC19} with an application to Massive MIMO. Introducing
multiple tokens in RPNs was also examined 
in~\cite{DBLP:journals/tcs/PhilippouP22} 
by allowing multiple  tokens of the same type to exist within a net 
following the collective interpretation and yielding a locally-controlled,
out-of-causal-order reversibility semantics. RPNs have been translated to 
Answer Set Programming (ASP), a declarative programming framework with 
competitive solvers~\cite{ASPtoRPNs}, and 
to bounded Coloured Petri Nets~\cite{RPNtoCPN,DBLP:conf/rc/BarylskaGMPPP22}.
\section{Reversing Petri Nets with Multiple tokens}

In our previous works we introduced Reversing Petri Nets, a net-based
formalism, which features individual tokens that can be connected
together via bonds~\cite{RPNscycles}. An assumption
of RPNs is that tokens are pairwise distinct. To relax this restriction,
subsequent work~\cite{DBLP:journals/tcs/PhilippouP22} introduced token 
multiplicity whereby a model may contain multiple tokens of the same type.
It was observed that the possibility
of firing a transition 
multiple times using different sets of tokens,  may introduce  
nondeterminism, also known as backward conflict, when going backwards. 
Furthermore, two approaches were identified to define reversible
semantics in the presence of such backward conflicts,
inspired by the individual token and the 
collective token interpretations~\cite{individual,ConfStruct}, defined
to reason about causality in Petri nets.
In the individual token approach, multiple tokens of the same type
residing in the same place are distinguished based on their causal path, 
whereas in the collective token interpretation 
they are not distinguished.
In~\cite{DBLP:journals/tcs/PhilippouP22} the model of RPNs
with multiple tokens was investigated under the collective token approach, 
yielding an out-of-causal-order form of reversibility.
In this work, we instead apply the individual token interpretation
to define a causal semantics, and we
establish that in fact the addition of multiple tokens does not add to
the expressiveness of the model, in that for any RPN with multiple
tokens there exists an equivalent RPN with only a single token of each type.
\begin{figure}[tb!]
	\centering
	\includegraphics[height=3cm]{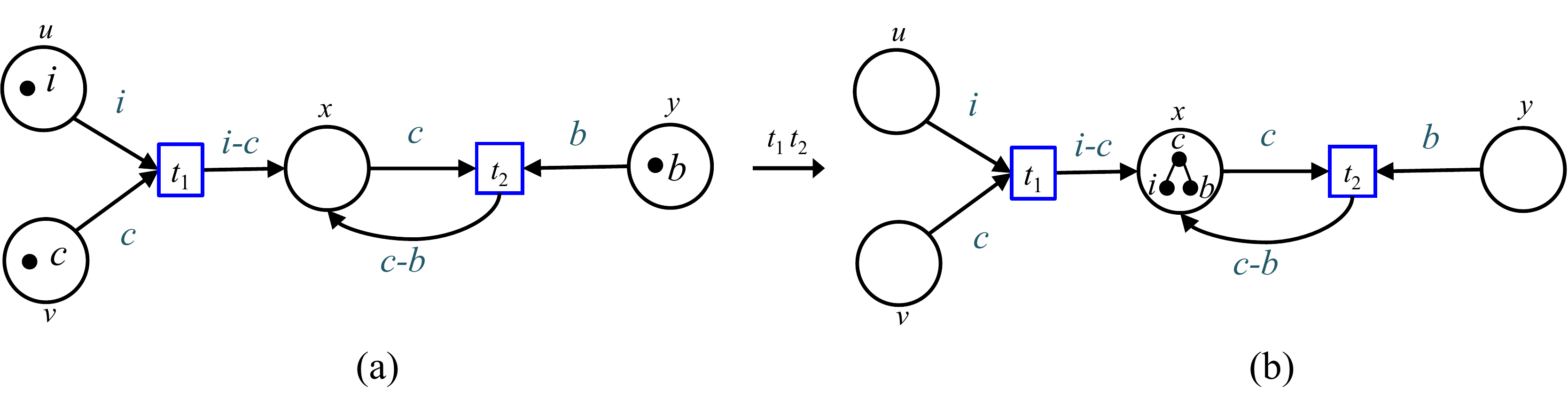}
	\caption{RPN example of a pen assembly/dissassembly}
	\label{RPN-pen}
\end{figure}

To appreciate the challenges induced through the introduction of multiple
tokens and the difference between
the individual and the collective token interpretations, let us consider
the example in Fig.~\ref{RPN-pen}(a). In this example we may see
an RPN model of an assembly/disassembly of a pen.
The product consists of the ink, the cup, and
the button of the pen, modelled by tokens $i$, $c$, and $b$, respectively.
We may observe
that transitions, in addition to transferring tokens between places,
have the capacity of creating bonds. Thus, the process of manufacturing the 
pen requires the ink to be fitted
inside the cup, modelled by the creation of the bond $i-c$ by
transition $t_1$ and, subsequently, the fitting of the button on
the cup to complete the  assembly, modelled as the creation
of the bond $c-b$ by transition $t_2$ (RPN in Fig.~\ref{RPN-pen}(b)).
The effect of reversing a transition in RPNs is to break the bonds created by the 
transition (if any) and returning the tokens/bonds from the outgoing places to
the incoming places of the transition. In~\cite{RPNs,RPNscycles} machinery
has been developed in order to model backtracking, causal, and 
out-of-causal-order reversibility for the model. In particular,
in the example of Fig.~1(b) reversing transition $t_2$
will result in the destruction of bond $c\bond b$ and the return of
token $b$ to place $y$.

Suppose we wish to extend the model of Fig.~1(a) for the assembly of
two pens. Given that  in RPNs tokens are unique, it would
be necessary to introduce three new and distinct
tokens and clone the transitions while renaming
their arcs to accommodate
for the names of the new tokens to be employed, resulting
in a considerable expansion of the model for
each new pen to be produced. Thus, a natural
extension of the formalism involves relaxing this restriction and
allowing multiple tokens of the same type to exist within a model.
To this effect consider the scenario of Fig.~2(a) 
presenting a system with an already assembled/sample pen in place
$x$ and two items of each of the ink, cup, and button components.

\begin{figure}[t]\label{MRPNpen}
\centering
\subfigure[]
    {\includegraphics[width=0.52\textwidth]{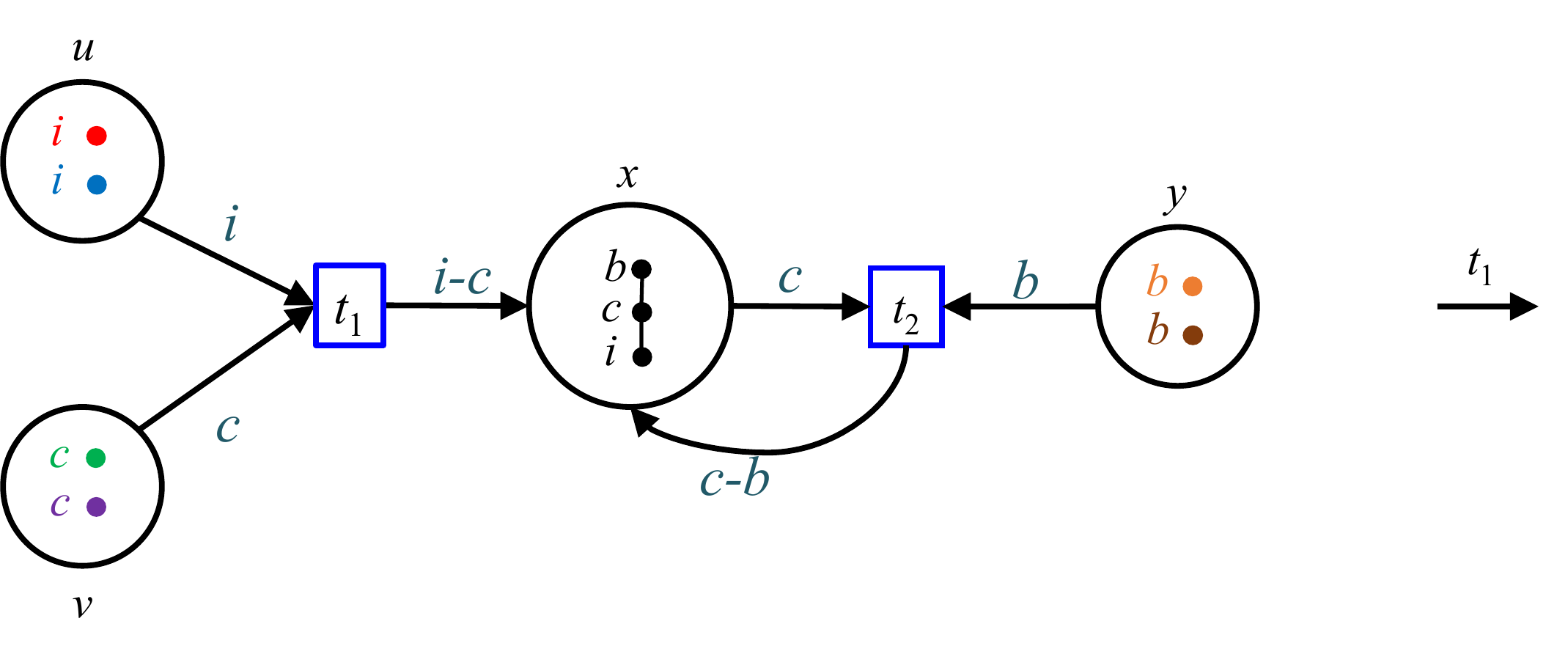}}\label{MRPN-pen1}
\subfigure[]
    {\includegraphics[width=0.47\textwidth]{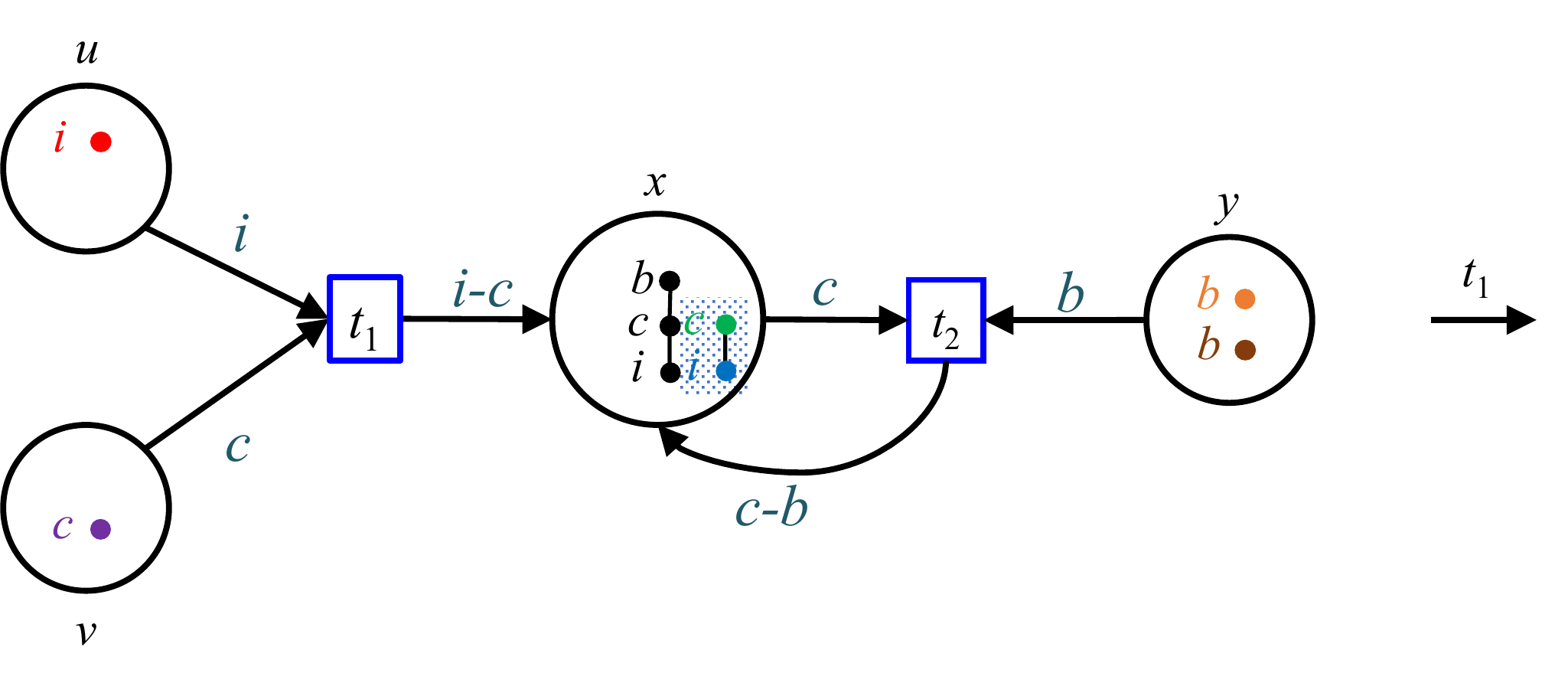}}\label{MRPN-pen2}
\subfigure[]
    {\includegraphics[width=0.55\textwidth]{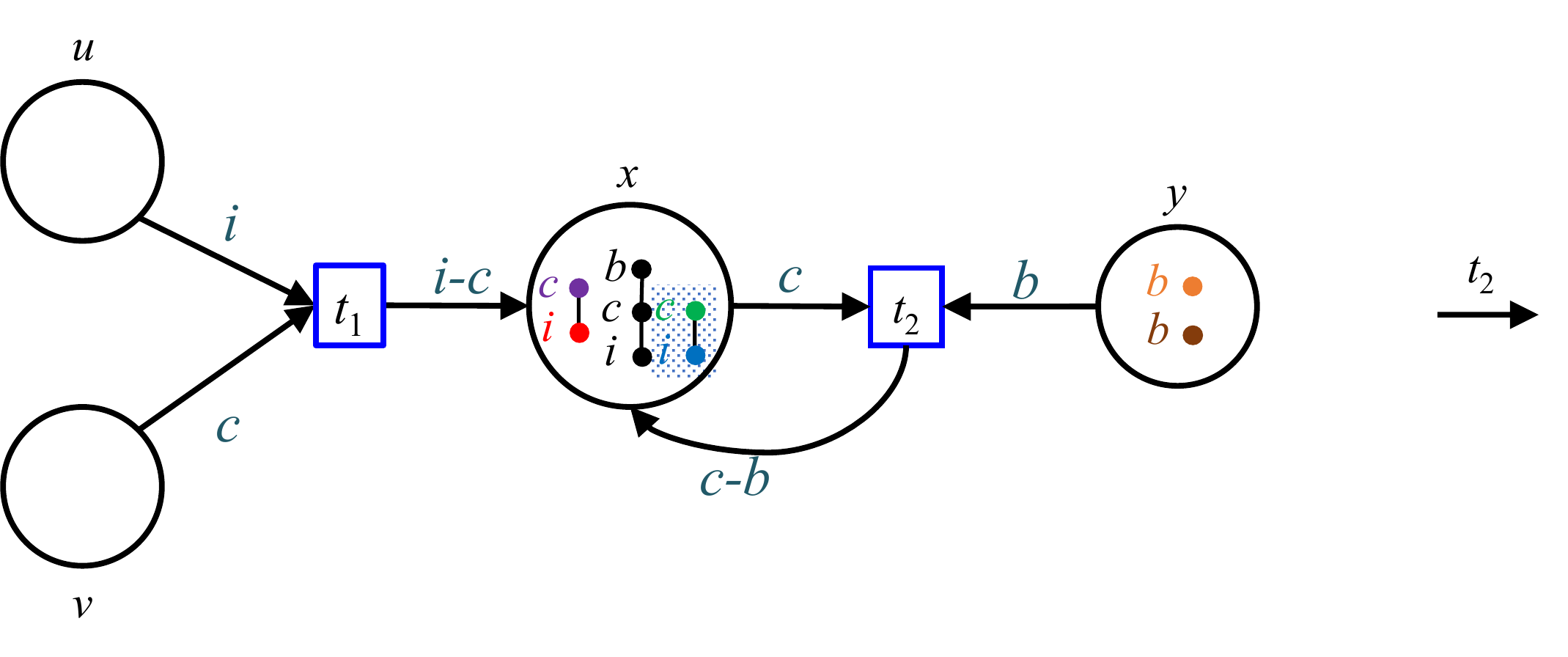}}\label{MRPN-pen3}
\subfigure[]
    {\includegraphics[width=0.44\textwidth]{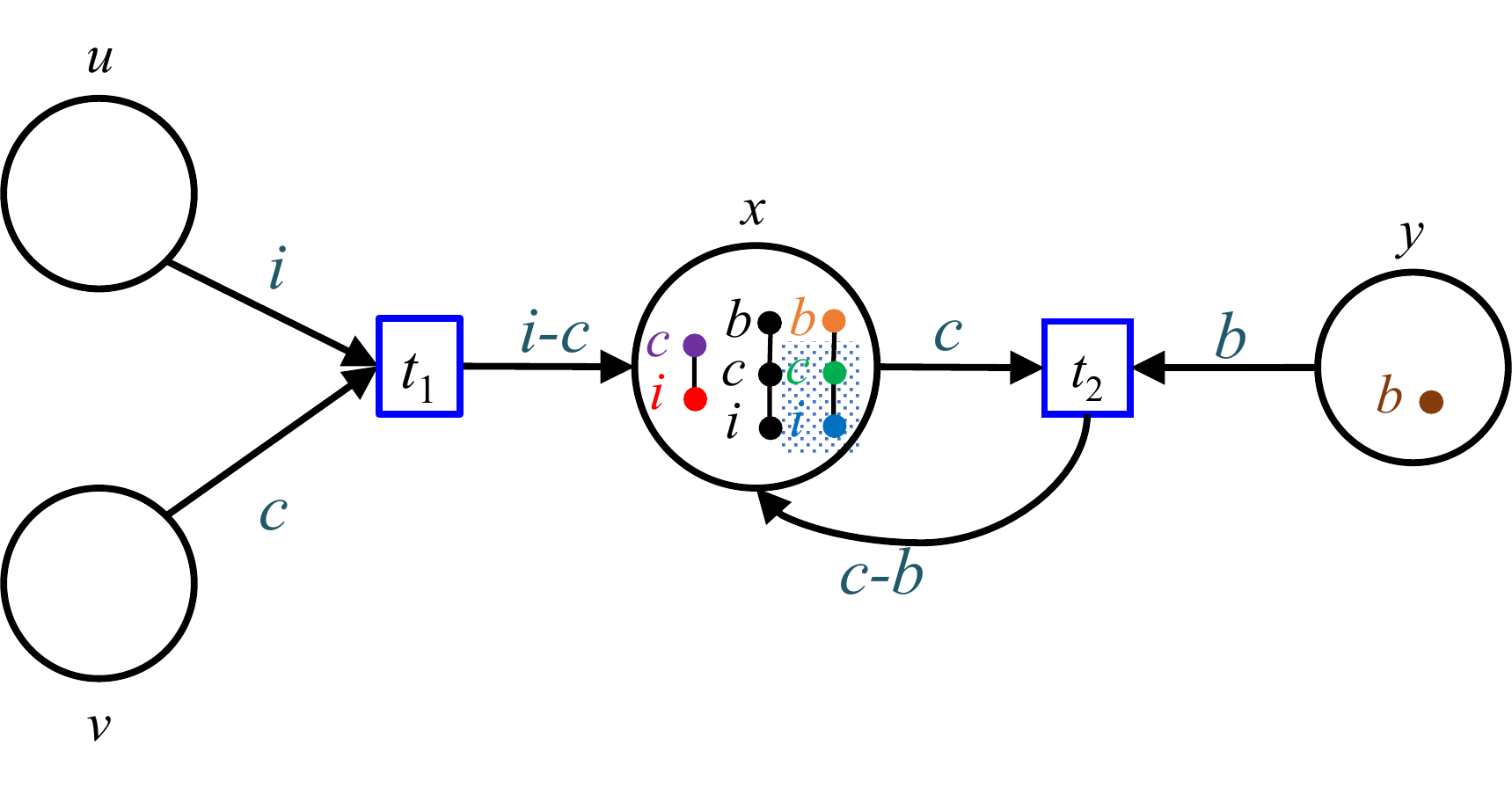}}\label{MRPN-pen4}
\caption{Executing transition $t_1$ in the net (a) may yield the net in (b). 
Different selections of tokens could have been made. In net (b) transition
$t_1$ is executed with the (only) available tokens leading to net (c), whereby
execution of $t_2$ with the component produced by the first execution
of $t_1$ yields net (d).}
\end{figure}

An issue arising in this new setting is that due to
the presence of multiple tokens of the same type, the phenomenon
of backwards nondeterminism occurs when transitions
are reversed. For instance, after execution of transition $t_1$ twice and $t_2$,
two assembled pens will exist in place $x$ and well as a component $i-c$,
as seen in Fig.~2(d). Suppose that in this state transition $t_1$ is reversed.
In the collective token interpretation, all instances of the bond $i-c$
are considered identical. As a result, any of these bonds could
be destroyed during the reversal of transition $t_1$.
However, in the individual token interpretation the various ink and cup tokens
are distinguished based on their causal path. Therefore, the first
execution of transition $t_1$ yielding the net in
Fig.~2(b) and involving the shaded component of tokens in
the figure, is considered to have caused the execution of transition $t_2$.
Given this causal relationships between the transitions, under a causal
reversibility semantics, the specific $i-c$ component should not be decomposed
until transition $t_2$ is reversed. Similarly, the pre-existing pen
should not be broken down into its parts as it
was not the created by any of the transitions. Instead, reversing transition
$t_1$ in the RPN of Fig. 2(d) should break the bond in the component
consisting the single bond $i-c$.
Note that this is compatible with the understanding that
disassembly of the product would not allow the
separation of the ink from the inside of the cup before the button
is removed, since this is enclosed within the pair
of the cup and the button. 

As a result we observe that following the individual token interpretation,
reversing a computation requires keeping track of past behavior -- in the
context of the example, distinguishing the tokens involving the pre-existing 
pen and the tokens used to fire each transition.
In the following sections we implement this approach for 
introducing multiple tokens and we study its
properties in the context of causal-order reversibility.
Furthermore, we establish a correspondence between
this model and RPNs with single tokens.

\section{Multi Reversing Petri Nets}

We present multi reversing Petri nets, an extension of RPNs with multiple
tokens of the same type
that allow transitions to be reversed following the individual token
interpretation. Formally, they are defined as follows:
\begin{definition}{\rm
A \emph{multi reversing Petri net} (MRPN) is a tuple $(P,T,\A,\A_V,\B,F)$ where:
	\begin{enumerate}
		\item $P$ is a finite set of \emph{places} and
		$T$ is a finite set of \emph{transitions}.
		\item $\A$ is a finite set of \emph{base} or \emph{token types} ranged 
		over by $A, B,\ldots$
		\item $\A_V$ is a finite set of \emph{token variables} ranged over by
		$a,b,\ldots$ We write 
		$\type(a)$ for the type
		of variable $a$ and assume that $\type(a) \in \A$ for all $a\in \A_V$.
		\item{$\B\subseteq \A\times \A$ is a finite set of undirected \emph{bond
		types} ranged over
		by $\beta,\gamma,\ldots$ We assume $\B$ to be a symmetric relation and we
		consider the elements $(A,B)$ and $(B,A)$ to refer to the same bond type,
		which we also denote by $A \bond B$.  Furthermore,
		we write $\B_V\subseteq \A_V\times \A_V$, assuming that $(a,b)$ and $(b,a)$
		represent the same bond, also denoted as $a-b$.}
		\item $F : (P\times T  \cup T \times P)\rightarrow {\cal P}( \A_V \cup 
		\B_V)$ 
		defines a set of directed labelled \emph{arcs} each associated with a 
		subset of $\A_V \cup \B_V$, where $(a,b)\in F(x,y)$ implies
		that $a,b\in F(x,y)$. Moreover, for all $t\in T$, $x,y\in P$, 
		$x\neq y$, $F(x,t)\cap F(y,t)=\emptyset$.
		\end{enumerate}
}\end{definition}

A multi reversing Petri net is built on the basis of a set of \emph{token
types}. Multiple occurrences of a token type, referred to as \emph{token
instances}, may exist in a net.
Tokens of the same type have identical 
capabilities on firing transitions and can participate only in transitions with
variables of the same type.

As standard in net-based frameworks, places and transitions are connected via  
labelled directed arcs. These labels are derived from $\A_V \cup \B_V$. They express the requirements and the effects of transitions based 
on the type of tokens consumed. Thus, collections of tokens corresponding to the 
same  types and connections as the variables on the labelled arc are able to 
participate in the transition.  
More precisely, if $F(x,t) = X\cup Y$, where $X\subseteq \A_V$, $Y\subseteq 
\B_V$, the firing of $t$ requires a distinct token instance
of type  $\type(a)$ for each $a\in X$, such that
the overall selection of tokens are connected together
satisfying the restrictions posed by $Y$. 
Similarly, if  $F(t,x) = X\cup Y$, where 
$X\subseteq \A_V$, $Y\subseteq  \B_V$,
this implies that during the forward execution of the transition
for each $a\in X$ a token instance  of type $\type(a)$ will be 
transmitted to place $x$ by the transition, in addition to  the bonds specified
by $Y$, some of which will be created as an effect of the transition.
We make the assumption that if $(a,b)\in Y$ then $a,b\in X$ and 
the same variable cannot be used on two incoming arcs of a 
transition.

We introduce the following notations. We write 
$\circ t =   \{x\in P\mid  F(x,t)\neq \emptyset\}$ and  
$ t\circ= \{x\in P\mid F(t,x)\neq \emptyset\}$
for the incoming and outgoing places of transition
$t$, respectively. Furthermore, we write
$\guard{t}  =   \bigcup_{x\in P} F(x,t)$ for the union of all labels on 
the incoming arcs of  transition $t$, and
$\effects{t}  =   \bigcup_{x\in P} F(t,x)$ for the union of all labels on the outgoing arcs of transition $t$.

We restrict our attention to well-formed MRPNs, which satisfy the conservation
property~\cite{conservative} in the sense that the number of 
tokens in a net remains constant during execution. In fact, as we will
prove in the sequel, in well-formed nets individual tokens are conserved.

\begin{definition}\label{multiwell-formed}{\rm 
	An MRPN  $(P,T,\A,\A_V,\B,F)$ is \emph{well-formed} if for 
	all $t\in T$:
	\begin{enumerate}
		\item $\A_V\cap \guard{t} = \A_V\cap \effects{t}$ and 
		\item
	 $ F(t,x)\cap F(t,y)=\emptyset$ for all $x,y\in P$, $x\neq y $. 
		\end{enumerate}
}\end{definition}
Thus, a well-formed MRPN satisfies (1) whenever a variable 
exists in
the incoming arcs of a transition then it also exists on its outgoing arcs, and 
vice versa, which
implies that transitions neither create nor erase tokens, and  (2) 
tokens/bonds cannot be cloned into more than one outgoing place.

In the context of token multiplicity,
a mechanism is needed in order to distinguish between token instances
with respect to their causal path.
For instance, consider the MRPN in 
Fig.~2(d). In this state, three connected components of tokens are positioned in
place $x$, where tokens of
the same type, e.g. the three  $c$ tokens  have distinct
connections and causal histories.
To capture this,
we distinguish between token instances, as follows:
\begin{definition}\label{recursive}{\rm	
		Given an MRPN $(P, T, \A, \A_V,\B,F)$ a \emph{token instance}  has the form
	$(A,i,xs)$ where $xs$ is a (possibly empty) list of triples $[(k_1,t_1,v_1), \ldots,(k_n,t_n, v_n)]$ with $n\geq 0$, where $i\geq 1$,
			$A\in \A$, and for all $i$,  $k_i\in \mathbb{N}$, $t_i\in T$,
			and $v_i \in  \{*\} \cup\A_V$.
We write $\A_I$ for the set of token instances ranged over by $A_1$, $A_2,\ldots$,
and we define the set of bond instances $\B_I$ by $\B_I=\A_I\times \A_I$.
Furthermore, given $A_i = (A,i,[(k_1,t_1,v_1),\ldots,(k_n,t_n,v_n)])$, we write 
\begin{eqnarray*}
\type(A_i) & = A& \\
A_i\da & = & (A,i)\\
\cpath{A_i} & = & [(t_1,v_1),\ldots,(t_n,v_n)]\\
\last{A_i} & = &  (k_n,t_n,v_n)\\
A_i + (k,t,v) & =&  (A,i,[(k_1,t_1,v_1),\ldots,(k_n,t_n,v_n),(k,t,v)])\\
\init{A_i} & = &  (A,i,[(k_1,t_1,v_1),\ldots,(k_{n-1},t_{n-1},v_{n-1})])
\end{eqnarray*}
}\end{definition}

The set of token instances $\A_I$  corresponds to the basic entities that
occur in a system. In the initial state of a net, tokens have the form
$(A,i,[])$ where $i$ is a unique identifier for the specific token instance
of type $A$.
As computation proceeds the tokens evolve to capture their causal
path. If a transition $t$ is executed in the forward direction,
with some token instance $(A,i,[(k_1,t_1,v_1),\ldots,(k_n,t_n,v_n)])$ substituted for
a variable $v$, then the token evolves to $(A,i,[(k_1,t_1,v_1),\ldots,
(k_n,t_n,v_n),(k,t,v)])$, 
where $k$ is an integer that characterizes the executed transition, as we will formally
define in the sequel.

In a graphical representation, tokens instances are indicated by $\bullet$ 
associated with their description, places by circles, transitions
by boxes, and bonds by lines  between tokens. Note that
token variables $a\in F(x,t) \cap \A_V$ with $\type(a) =A$ are 
denoted by $a:A$ over the corresponding arc $F(x,t)$.  An example of 
an MRPN can be seen in Fig. 3. In this example, we have $\A=\{I,C,B\}$, 
$\A_V=\{i,c,b\}$, and the set of token instances in the specific state are 
$\{(I,i,[]), (B,i,[]),(C,i,[])\mid i\in\{1,2,3\}\}$.

\begin{figure}[tb!]
	\centering
	\includegraphics[height=4cm]{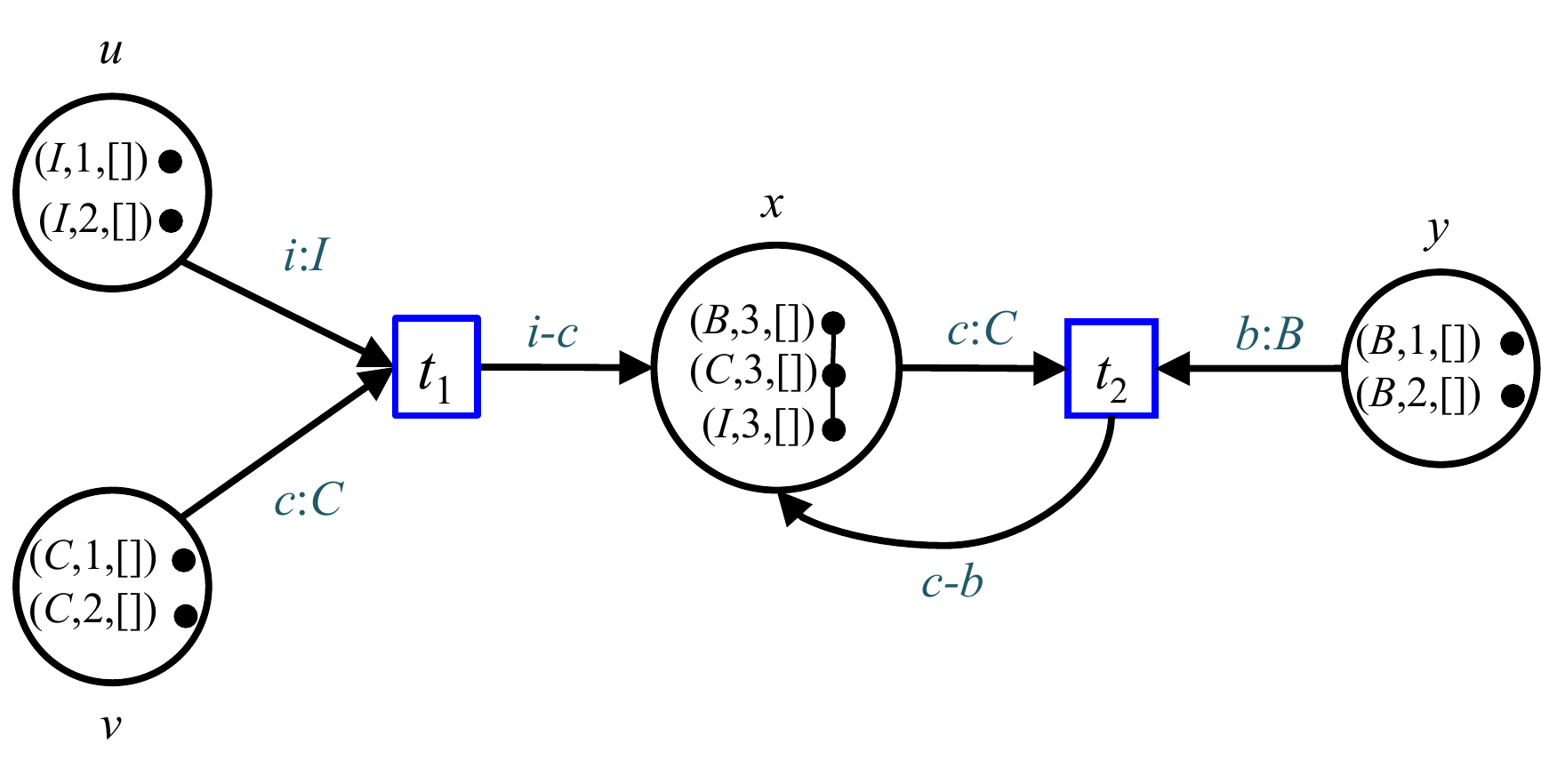}
	\caption{The net of Fig. 2(a) presented as an MRPN.}
	\label{RPNI-pen}
\end{figure}

As with RPNs the association of token/bond instances to places is called a \emph{marking}  such that 
$M: P\rightarrow 2^{\A_I\cup \B_I}$, where we assume that if $(A_i,B_i)\in M(x)$ then $A_i, B_i\in M(x)$. 
In addition, we employ the notion of a \emph{history}, which~assigns a memory to each
transition $H : T\rightarrow 2^\mathbb{N}$. 
Intuitively, a history of $H(t) = \emptyset$ for some $t \in T$ captures that the transition has not taken place, or 
every execution of it has been reversed, and a history
such that $ k \in H(t)$, captures that the transition had a firing
with identifier $k$ that was not reversed.
Note that $|H(t)|>1$ may
arise due to cycles but also due to the consecutive execution of the transition by different token
instances. A pair of a marking and a history, $\state{M}{H}$, describes a \emph{state} of an MRPN 
with $\state{M_0}{H_0}$ the initial state, where $H_0(t) = \emptyset$ for all $t\in T$ and if
$A_i \in M_0(x)$, $x\in P$, then $A_i=(A,i,[])$, 
and $A_i\in M_0(y)$ implies that $x=y$.

Finally, we define $\connected(A_i,W)$, where $A_i\in \A_I$ and $W\subseteq {\A_I\cup \B_I}$,
to be the tokens connected
to $A_i$  as well as the bonds creating these connections according to 
set $W$:
\begin{eqnarray*}
\connected(A_i,W)&=&(\{A_i\} \cap W)\\
        &\cup&\{x \mid \exists  w \mbox{ s.t. }  \paths(A_i,w,W), (B_i,C_i)\in w, x\in\{(B_i,C_i),B_i,C_i\} \}
\end{eqnarray*}
where $\paths(A_i,w,W)$ if $w=\langle\beta_1,\ldots,\beta_n\rangle$, and for all $1\leq i \leq n$, $\beta_i=(x_{i-1},x_i)\in W\cap \B_I$, $x_i\in W\cap \A_I$, and $x_0 = A_i$.
For example, consider the net in Fig.~3 and let $W$ represent the
set of token and bond instances in place $x$. Then, $\connected((I,3,[]),W)=\{(I_3, C_3, B_3, (I_3,C_3), (C_3,B_3)\}$, where $I_3=(I,3,[])$, $B_3=(B,3,[])$, and $C_3=(C,3,[])$.

\subsection{Forward Execution}
During the forward execution of a transition in an MRPN, 
a set of token and bond instances, as specified by the incoming arcs of the transition, are selected and
moved to the outgoing places of the transition, possibly
forming and/or destroying bonds. Precisely, for a transition $t$
we define $\effectp{t}$ to be the bonds 
that occur on its outgoing arcs but not the incoming ones and by $\effectm{t}$  
the bonds that occur in the incoming arcs but not the outgoing 
ones: 
\[
\effectp{t} = \effects{t} - \guard{t}\hspace{1cm}
\effectm{t} = \guard{t} -\effects{t} 
\]

Due to the presence of multiple instances of the same token
type, it is possible that different token instances are selected 
during the transition's execution.  To enable such a selection
of tokens  we define the following:
\begin{definition}{\rm \ 
An injective function ${\cal W}: V\rightarrow \A_I$, where 
$V\subseteq \A_V$, is called a {\em type-respecting assignment} if for
all $a\in V$, if ${\cal W}(a) = A_i$ then $\type(a) = \type(A_i)$.
}\end{definition}

We extend the above notation and write $\W(a,b)$ for 
$(\W(a),\W(b))$ and, given a set $L\subseteq \A_V\cup \B_V$, 
we write $\W(L) = \{\W(x) \mid x\in L\}$.

Based on the above we define the following:
%

\begin{definition}\label{multifenabled}{\rm
	Given an MRPN $(P,T, \A, \A_V, \B, F)$, a state $\state{M}{H}$, 
	and a transition $t$, we say that $t$ is 
	\emph{forward-enabled} in $\state{M}{H}$  if there exists a
	type-respecting assignment ${\cal S}:\guard{t}\cap 
	\A_V\rightarrow \A_I$ such that:
	\begin{enumerate}
		\item $\S(F(x,t))\subseteq M(x)$ for all $x\in \circ t$. 
		\item If $a,b \in F(x,t)$ for some $x\in \circ t$ and $(a,b)\in \effectp{t}$, then $\S(a,b) \not\in
	    M(x)$.
	    \item If $a \in F(t,y_1)$ and $b \in F(t,y_2)$, for some $y_1,y_2\in t\circ$,
		$y_1\neq y_2$, then 
		$\connected(\S(a),
		\after(t,\S,M))
		\neq \connected(\S(b),\after(t,\S,M))$.
		\end{enumerate}
where $ \after(t,\S, M)=(\bigcup_{x\in \circ t}M(x) \cup \S(\effectp{t} ))
-\S(\effectm{t})$.

}\end{definition}

Thus, $t$ is forward-enabled in state $\state{M}{H}$ if there exists a 
type-respecting assignment $\S$ of token instances 
to the variables on the incoming edges of $t$, which we will refer
to as a {\em forward-enabling assignment} of $t$, such that (1)
the token instances and bonds required by
the transition's incoming edges, according to $\S$,
are available from the appropriate input
places,
(2) if the selected token instances to be transferred
by the transition are to be bonded together by the transition
then they should not be already bonded in an 
incoming place of the transition (thus the bonds that occur only on the 
outgoing arcs of a transition are
the bonds being created by the transition), and  (3) 
if two token instances are transferred by a transition to different
outgoing places then these tokens should not be connected. This
is to ensure that connected components are not cloned. 
Note that $\after(t,\S,M)=(\bigcup_{x\in \circ t}M(x) \cup \S(\effectp{t})) 
-S(\effectm{t})$ denotes the set of token  and bond instances
that occur in the incoming places of $t$ ($\bigcup_{x\in \circ t}M(x)$),
including the new bond instances
created by $t$ ($\S(\effectp{t})$), and removing the bonds destroyed
by it ($\S(\effectm{t})$). 
Intuitively, $\after(t,\S,M)$ contains the components that are
moved forward by the transition.

To execute a transition $t$ according to an enabling assignment $\S$, 
the selected token instances along with their connected components
are relocated to the outgoing places of the transition as specified
by the outgoing arcs, with bonds created and destroyed accordingly. 
 An additional effect is
the update of the affected token and bond instances to capture the 
executed transition in their causal path.  To capture this update
we define where $k$ is an integer associated with the specific
transition instance:
\[
		\hspace{1.5 cm}
		A_i \oplus (\S,t,k) = \left\{
		\begin{array}{ll}
		A_i+(k,t,a)\hspace{1.5cm} \textrm{ if } \S(a)=A_i  \; \\
		A_i+(k,t,*)\hspace{1.5cm} \textrm{ if } \S^{-1}(A_i)=\bot 
		\end{array}
		\right.			\]
Note that $A_i$ may not belong to the range of $\S$, i.e. $\S^{-1}(A_i)=\bot$, 
if $A_i$ was not specifically selected to instantiate a variable in
$\guard{t}$ but, nonetheless, belonged to a connected component
transferred by the transition. This is recorded in the causal path
of the token instance via the triple $(k,t,*)$.
Moreover, we write $(A_i,B_j)\oplus(\S,t,k)$ for $(A_i\oplus(\S,t,k),B_j\oplus(\S,t,k))$ and,
given $L\subseteq \A_I\cup \B_I$, we write $L\oplus(\S,t,k)=\{x\oplus(\S,t,k)\mid x\in L\}$. Finally,
the history of the executed transition is updated to include
the next unused integer.
Given the above we define:
\begin{definition}{\rm \label{multiforward}
		Given an MRPN $(P,T, \A, \A_V, \B, F)$, a state $\langle M, H\rangle$, a transition $t$ that is enabled in state $\langle M, H\rangle$,  and an enabling assignment $\S$,
we write $\state{M}{H}
		\trans{(t,S)} \state{M'}{H'}$
		where for all $x\in P$:
		\begin{eqnarray*}
		M'(x) & = & 
		 (M(x)-\bigcup_{a\in F(x,t)}\connected(\S(a),M(x)))\\
	   &\cup& \;\bigcup_{a\in F(t,x)}\connected(\S(a),\after(t,\S,M)) \oplus(\S,t,k)
	   \end{eqnarray*}
		where 
		$k=max(\{0\} \cup H(t))+1$ and
		\[
		\hspace{1.5 cm}
		H'(t') = \left\{
		\begin{array}{ll}
		H(t') \cup  \{k\}, \hspace{1.4cm} 
		\textrm{ if } t' = t  \; \\
		H(t'), \hspace{2.33cm} \textrm{ otherwise }
		\end{array}
		\right.			\]
}\end{definition}

\begin{figure}[t]
	\begin{center}
		\textbf{}
		\includegraphics[height=3.2cm]{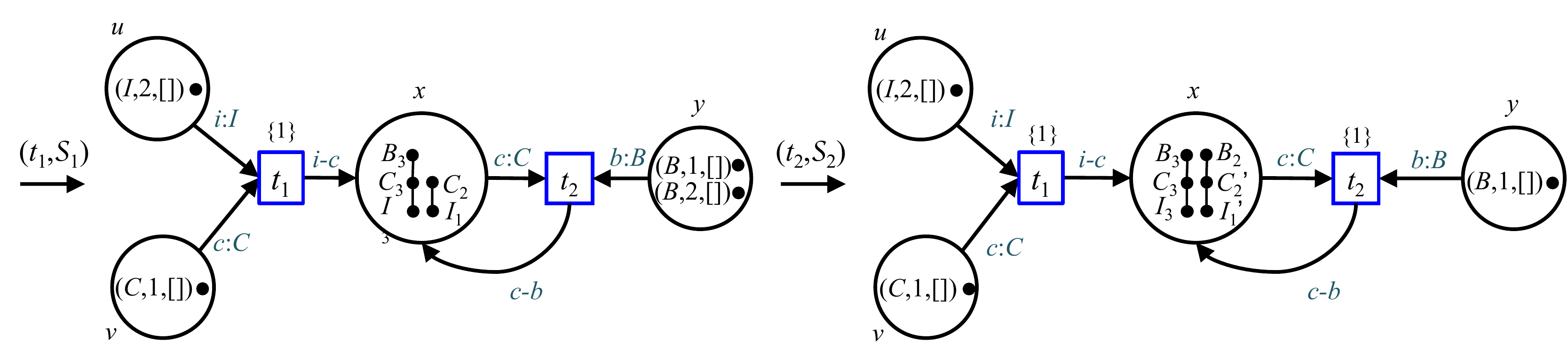}
	\end{center}
	\label{FPen}
	\caption{The effect of executing  $t_1$ and $t_2$ in the net of Fig. 3, where  $B_3=(B,3,[])$, $C_3=(C,3,[])$, $I_3=(I,3,[])$, $I_1=(I,1,[(1,t_1,i)])$, $C_2=(C,2,[(1,t_1,c)]$,  $I_1'=(I,1,[(1,t_1,i),(1,t_2,*)])$, $C_2'=(C,2,[(1,t_1,c),(1,t_2,c)])$, and $B_2=(B,2,[(1,t_2,b)])$.}
\end{figure}

Fig.~4 shows the result of consecutively firing transitions $t_1$ and $t_2$ from the MRPN 
in Fig. 3 with enabling assignments 
$\S_1$, where $\S_1(i) = (I,1,[])$, $\S_1(c) = (C,2,[])$, and $\S_2$,
where $\S_2(b) = (B,2,[])$, $\S_2(c) = (C,2,[(1,t_1,c)])$. We 
note the non-empty histories of the transitions depicted in the graphical representation,
as well as the updates in the causal paths of the tokens.

\subsection{Causal-order Reversing}
We now move on to consider causal-order reversibility for MRPNs.
In this form of reversibility, a transition can be reversed only
if all its  effects (if any), i.e. transitions that it has caused, have already been reversed. As
argued in~\cite{RPNscycles}, two transition occurrences are causally 
dependent, if a token produced by the one  was subsequently used to
fire the other.
Since token instances in MPRNs are associated with their causal path, we
are able to identify the transitions that each token has participated in by 
observing its memory. Furthermore, if $\last{A_i} = (k,t,a)$ then the last transition that
the token instance $A_i$ has participated in was transition $t$ and specifically
its occurrence with history $k$.

Based on this observation,
a transition occurrence  $t$ can be reversed in a certain state if the token/bonds instances
it has employed have not engaged in any further transitions.
Thus, we define  causal reverse enabledness as follows.
\begin{definition}\label{multico-enabled}{\rm
	Consider an MRPN $(P,T, \A, \A_V, \B, F)$, a state $\state{M}{H}$, and a transition $t$. 
		We say that 
		$t$ is \emph{co-enabled} in $\state{M}{H}$ if 
		there exists a type-respecting 
		assignment $\R:\effects{t}\cap \A_V\rightarrow  \A_I$ such that:
		\begin{enumerate} 
			\item  $\R(F(t,x))\subseteq M(x)$ for all $x\in t\circ$, and
			\item there exists $k\in H(t)$ such that for all $(A,i,xs)\in \bigcup_{x\in P}M(x)$ with $(k,t,b)\in xs$
			for some $b$, $(k,t,b)=\last{A_i}$.
		\end{enumerate}
We refer to $\R$ as the $co$-reversal enabling assignment for the $k^{th}$ occurrence of $t$.
}\end{definition}

Thus, a transition  $t$ is $co$-enabled in $\state{M}{H}$ for a specific
occurrence $k$ if  
there exists a type-respecting assignment of token instances on
the variables of the outgoing arcs of the transition, which gives
rise to a set of token and bond instances that are available in the
relevant out-places and, additionally, these token/bond instances 
were last employed for the firing of the specific occurrence of
the transition.

To implement the reversal of a transition $t$ according to a $co$-reversal
enabling assignment $\R$, the selected token instances are relocated from  the outgoing
places of $t$ to its incoming places, 
with bonds created and destroyed accordingly. 
The occurrence of the reversed transition is removed from its history.

\begin{definition}{\rm \label{multicausal}
		Given an MRPN $(P,T, \A, \A_V, \B, F)$, a state $\langle M, H\rangle$,  a transition $t$ 
		that is $co$-enabled with $co$-reversal
		enabling assignment $\R$ for the $k^{th}$ occurrence of $t$, 
		we write $\state{M}{H}
		\ctrans{(t,\R)} \state{M'}{H'}$
		where 
		for all $x\in P$:
		\begin{eqnarray*}	
			M'(x) &=&  (M(x)-\bigcup_{a\in F(t,x)}\connected(\R(a),M(x))) \\
		&\cup& \;\bigcup_{a\in F(x,t)}\init{\connected(\R(a),\before(t,\R,M))}
	   \end{eqnarray*}
	  and
		\[
		H'(t') = \left\{
		\begin{array}{ll}
		H(t') - \{k\}, \hspace{.5cm} \textrm{ if } t' = t  \; \\
		H(t'), \hspace{1.38cm}  \;\textrm{ otherwise }
		\end{array}
		\right.			\]	
			where $\before(t,\R, M)=(\bigcup_{x\in t\circ}M(x) \cup \R(\effectm{t} ))-\R(\effectp{t})$.
}\end{definition}  
\begin{figure}[t]
	\begin{center}
		\textbf{}
		\includegraphics[height=3.2cm]{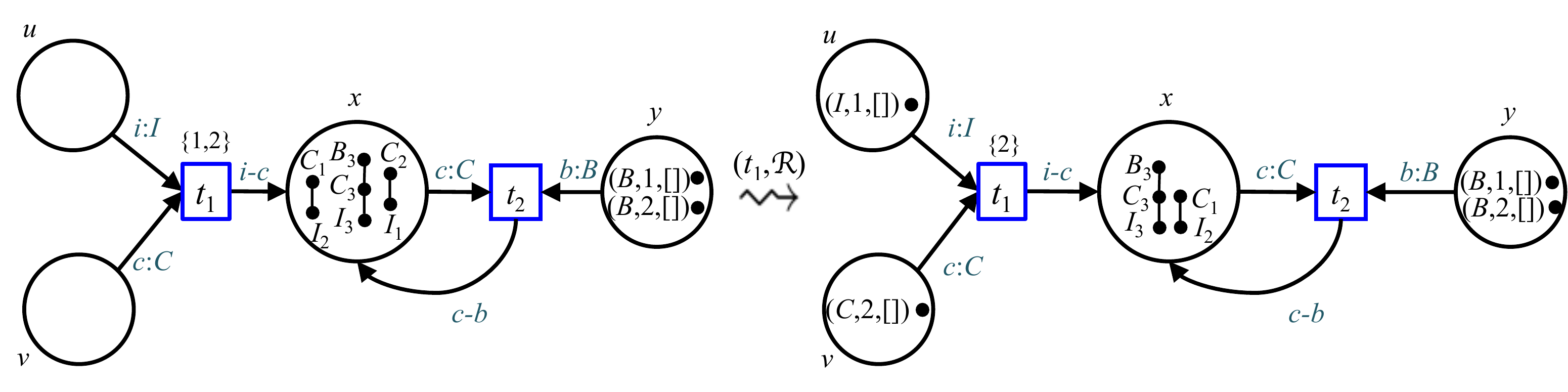}
	\end{center}
	\label{CRPen}
	\caption{The effect of reversing transition $t_1$ with enabling assignment
	$\R(i)=I_1$, $\R(c)=C_2$, in a state following the execution of
	$t_1$ twice from the net in Fig. 3, first with enabling assignment
	$\S_1(i)=(I,1,[])$, $\S_1(c) = (C,2,[])$, and next with enabling assignment
	$\S_2(i)=(I,2,[])$, $\S_2(c) = (C,1,[])$ where we write $I_1=(I,1,[(1,t_1,i])$, $I_2=(I,2,[(2,t_1,i])$, 
	$C_1=(C,1,[(2,t_1,c])$, and $C_2=(C,2,[(1,t_1,c])$.}
\end{figure}
In Fig.~5  we may observe the causal-order reversal of 
transition $t_1$. We note that the history information of the affected 
components is updated by removing the occurrence of the reversed transition
and the history information of transition $t_1$ reflects that 
occurrence with identifier $1$ has been reversed.

Let us now consider executions of both forward and backward moves and 
write $\fctrans{}$ for $\trans{}\cup\ctrans{}$. We define the
reachable states of an MRPN as follows.
\begin{definition}{\rm\
Given an MRPN $(P,T,\A,\A_V,\B,F)$ and an initial state 
$\langle M_0, H_0\rangle$ we say that state 
$\langle M,H\rangle$ is \emph{reachable}, if there exist
$\langle M_i, H_i\rangle$, $i\leq n$ for some $n\geq 0$, such that
$\state{M_0}{H_0} \fctrans{(t_1,\W_1)}\state{M_1}{H_1} \fctrans{(t_2,\W_2)}\ldots \fctrans{(t_n,\W_n)}\state{M_n}{H_n} =\state{M}{H}$.
}\end{definition}
Furthermore, given a type $A$, an integer $i$, and a marking $M$,
we write $\num{A,i,M}$  for the number of 
token instances of the form
$(A,i,xs)$ in $M$, defined by $\num{A,i,M}=|\{(x,A_i) \mid \exists x\in P, A_i\in M (x), A_i\da=(A,i)\}|$. 
Similarly, for a bond instance $\beta_i\in \B_I$, we define
$\num{\beta_i,M}=|\{x\in P 
\mid \beta_i\in M(x)\}|$.
The following result confirms that in an execution beginning in the initial
state of an MRPN, token instances are 
preserved, at most one bond instance may occur at any time, and  a bond instance may be created/destroyed
during a forward/reverse execution of a transition that features the bond as its effect.

\begin{proposition}\label{Prop}{\rm Given an MRPN $(P,T,\A,\A_V,\B,F)$,
a reachable state $\state{M}{H}$, and
a transition firing
$\state{M}{H} \fctrans{(t,\W)}\state{M'}{H'}$, the following hold:
\begin{enumerate}
\item For all $A$, $i$,   
    $\num{A,i,M'} = \num{A,i,M}$=1.
\item For all $\beta_i\in \B_I$, 
\begin{enumerate}
\item $0 \leq \num{\beta_i,M'} \leq 1$,
\item if $t$ is executed in the forward direction with
forward enabling assignment $\S$ and $\beta_i\in \S(\effectp{t})$ then 
$\num{\beta_i,M'} = 1$; if instead $\beta_i\in\S(\effectm{t})$ then 
$\num{\beta_i,M'} = 0$, otherwise 
$\num{\beta_i,M} = \num{\beta_i,M'}$.
\item if $\,t\,$ is executed in the reverse direction with reverse enabling assignment
$\;\R\;$ and $\beta_i\,\in$ $\R(\effectp{t})$ then $\num{\beta_i,M'} = 0$; 
if instead $\beta_i\in\R(\effectm{t})$ then $\num{\beta_i,M'} = 1$, 
otherwise $\num{\beta_i,M}=\num{\beta_i,M'}$.
\end{enumerate}
\end{enumerate}
\emph{Proof:}
The proof  follows by induction on the length of the execution
	reaching state $\state{M}{H}$. If this is the initial state the 
	result (i.e. clauses 1 and 2(a)) follows by our assumption on the initial 
	state. For the induction step, let us assume that $\state{M}{H}$ satisfies
	the conditions of the proposition.

Let us begin with clause (1) and suppose $\fctrans{(t,\W)} = \trans{(t,\S)}$, 
where $\S$ is the forward-enabling
assignment for the transition, and let $A_i=(A,i,xs)\in \A_I$. Two cases exist:
\begin{enumerate}
\item $A_i\in \connected(B_j,M(x))$ for some $B_j$, $\S(a) = B_j$, $a\in F(x,t)$. 
Note that $x$ is unique by the
assumption that $\num{A,i,M} = 1$. 
To discern the location of $A_i$ in
$M'$ two cases exist.
\begin{itemize}
    \item 
Suppose $A_i\in\connected(B_j,\after(t,\S,M))$.
We observe that,  by  Definition~\ref{multiwell-formed}(1),
$a\in \effects{t}$. Thus, there exists $y\in t\circ$, such that $a\in F(t,y)$. Note that this $y$
is unique by Definition~\ref{multiwell-formed}(2).  As a result, by Definition~\ref{multiforward}, 
$\connected(B_j,\after(t,\S,M))\subseteq M'(y)$, which implies that $A_i\in M'(y)$. 
\item Suppose $A_i\not\in \connected(B_j,\after(t,\S,M'))$ and
consider $w=\langle (A_{i_1},A_{i_2}),\ldots,$ $(A_{i_n},B_j)\rangle$, $A_i=A_{i_1}$, $n\geq 1$,
such that $\paths(A_i,w,M(x))$. Since $A_i\not\in \connected(B_j,\after(t,\S,M'))$ 
it must be that for some $k$,
$(A_{i_{k-1}},A_{i_{k}})\in \S(\effectm{t})$ and $A_i\in\connected(A_{i_k},M(x)-\S(\effectm{t}))$. Using the
same argument as in the previous case for $A_{i_k}$ instead of
$B_j$, we may conclude that $A_i\in M(y)$ such that $\S(b)=A_{i_k}$ and $b\in F(t,y)$.
\end{itemize}
Now suppose that $A_i\in \connected(C_k,\after(t,\S,M))$, $C_k = \S(b)$ for some  $b\neq a$, 
$b\in F(t,y')$. 
Then 
it must be that $y = y'$. As a result, we have that $\num{A,i,M'} = \num{A,i,M} = 1$ and 
the result follows.
\item $A_i\not\in \connected(\S(b),M(x))$ for all $b\in F(x,t)$, $x\in P$. This implies that 
$\{x\in P\mid A_i\in M'(x)\} = \{x\in P\mid A_i\in M(x)\}$ and the result follows.
\end{enumerate}

Now suppose $\fctrans{(t,\W)} = \rtrans{(t,\R)}$ where 
 $\R$ is the reverse-enabling
assignment of the transition. Consider $A_i=(A,i,xs)\in \A_I$. Two cases exist:
\begin{enumerate}
\item $A_i\in \connected(B_j,M(x))$ for some $B_j$, $\R(a) = B_j$, $a\in F(t,x)$. 
Note that $x$ is unique by the
assumption that $\num{A,i,M} = 1$. 
To discern the location of $A_i$ in
$M'$ two cases exist.
\begin{itemize}
    \item Suppose $A_i\in\connected(B_j,\before(t,\R,M'))$.
We observe that, 
by  Definition~\ref{multiwell-formed}(1),
$a\in \guard{t}$. Thus, there exists $y\in \circ t$, such that $a\in F(y,t)$. Note that this $y$
is unique by Definition~\ref{multiwell-formed}(3).  As a result, by Definition~\ref{multicausal}, 
\[M'(y) = M(x)-\bigcup_{a\in F(t,x)}\connected(\R(a),M(x)))
		\cup \;\bigcup_{a\in F(x,t)}\init{\connected(\R(a),\before(t,\R,M))}\]
		Since
$a\in F(y,t)\cap F(t,x)$, $A_i\in \connected(\R(a),M(x)\cup F(y,t))$, which implies that $a\in M'(y)$. 
\item Suppose $A_i\not\in \connected(B_j,\before(t,\R,M'))$ and
consider $w=\langle (A_{i_1},A_{i_2}),\ldots,$ $(A_{i_n},B_j)\rangle$, $A_i=A_{i_1}$, $n\geq 1$, 
such that $\paths(A_i,w,M(x))$. Since $A_i\not\in \connected(B_j,$ $\before(t,\R,M'))$ it must be that for some $k$,
$(A_{i_{k-1}},A_{i_k})\in $  $\R(\effectp{t})$ and $A_i\in\connected(A_{i_k},M(x)-\R(\effectp{t}))$. Using the
same argument as in the previous case for $A_{i_k}$ instead of
$B_j$, we may conclude that $A_i\in M(y)$ such that $\S(b)=A_{i_k}$ and $b\in F(y,t)$.
\end{itemize}
Now suppose that $A_i\in \connected(C_k,\before(t,\R,M))$, $C_k = \R(b)$ for some  $a\neq b$, 
$b\in F(y',t)$. 
Then 
it must be that $y = y'$. As a result, we have that $\{z\in P\mid A_i\in M'(z)\}
= \{y\}$ and  the result follows.
\item $A_i\not\in \connected(\R(a),M(x))$ for all $a\in F(t,x)$, $x\in P$. This 
implies that $\{x\in P\mid A_i\in M'(x)\} = \{x\in P\mid A_i\in M(x)\}$ and the 
result follows.
\end{enumerate}
The proof of clause 2 follows similar arguments.
\proofend}
\end{proposition}

We may now proceed to establish the causal consistency of our semantics. 
We begin with defining when two states of an MRPN are considered to be 
causally equivalent. Intuitively, states $\state{M}{H}$ and
$\state{M'}{H'}$ are causally equivalent
whenever the executions that have led
to them contain the same
causal paths. Note that these causal paths refer to different independent
threads of computation, possibly executed through 
different interleavings in the executions leading to $\state{M}{H}$ 
and $\state{M'}{H'}$. In our setting, we can enunciate this requirement
by observing the causal histories of token instances and requiring 
that for each token instance of some type $A$ in one of the two states
there is a token instance of the same type that has participated in
the exact same sequence of transitions in the other state:
\begin{definition}\label{multieqmarkings}{\rm Consider MRPN 
$(P,T, \A, \A_V, \B, F)$ and reachable states $\state{M}{H}$,
$\state{M'}{H'}$. Then the states are \emph{causally equivalent},
denoted by $\state{M}{H}\asymp\state{M'}{H'}$,
if for each $x\in P$, $A_i\in M(x)$ there exists $A_j\in M'(x)$
with $\cpath(A_i)= \cpath(A_j)$, and vice versa.
}\end{definition} 
We may now establish the Loop Lemma for our model.
\begin{lemma}[Loop]\label{multiloopc}{\rm 
		For any forward transition $\state{M}{H}\trans{(t,\S)}\state{M'}{H'}$ there exists a backward transition
		$\state{M'}{H'} \ctrans{(t,\R)} \state{M}{H}$ and for any backward transition $\state{M}{H} \ctrans{(t,\R)} \state{M'}{H'}$ there exists a forward transition $\state{M'}{H'}\trans{(t,\S)}\state{M''}{H''}$ where $\state{M}{H}\asymp \state{M''}{H''}$.
}\end{lemma}
\emph{Proof:}
Suppose $\state{M}{H}\trans{(t,\S)}\state{M'}{H'}$. Then $t$ is clearly reverse-enabled 
in $\state{M'}{H'}$ with reverse-enabling assignment $\R$ such
that if $\S(a)=(A,i,xs)$, then $\R(a)=(A,i,xs+(t,k,a))$, where
$k$ is the maximum element of $H(t)$. 
Furthermore, $\state{M'}{H'} \ctrans{t,\R} \state{M''}{H''}$
where $H'' = H$. In addition, all token and bond instances involved in transition
$t$ (except those in $\effectp{t}$) will be returned from the outgoing places of
transition $t$ back to its incoming places. 
At the same time, all destroyed bonds (those in $\effectm{p}$) will be re-formed,
according to Proposition~\ref{Prop}.
Specifically, for all $A_i\in \A_I$, it
is easy to see by the definition of $\ctrans{}$  that $A_i\in M''(x)$ if and only if 
$A_i\in M(x)$. Similarly,  for all $\beta_i\in \B_I$, $\beta_i\in M''(x)$ if and
only if $\beta_i\in M(x)$. The opposite direction can be argued similarly,
with the distinction that when a transition is executed immediately following
its reversal, it is possible that the transition instance is assigned a different
key, thus giving rise to a state $\state{M''}{H''}$ distinct but causally
equivalent to $\state{M}{H}$.
\proofend

We now proceed to define 
some auxiliary notions. Given a transition $\state{M}{H}\fctrans{(t,\W)}\state{M'}{H'}$,
we say that the \emph{action} of the transition is
$(t,\W)$ if $\state{M}{H}\trans{(t,\W)}\state{M'}{H'}$ and $(\underline{t},\W)$ 
if $\state{M}{H}\ctrans{(t,\W)}\state{M'}{H'}$
and we may write $\state{M}{H}\fctrans{(\underline{t},\W)}\state{M'}{H'}$. 
We write $Act_N$ for the set of all actions in an MRPN $N$.
We use $\alpha$ to
range over $\{t,\underline{t} \mid t\in T\}$ and write 
$\underline{\underline{t}} = t$. Given an execution 
$\state{M_0}{H_0}\fctrans{(\alpha_1,\W_1)}\ldots 
\fctrans{(\alpha_n,\W_n)}\state{M_n}{H_n}$, we say that the \emph{trace} 
of the execution is
$\sigma=\langle (\alpha_1,\W_1),(\alpha_2,\W_2),\ldots,(\alpha_n,\W_n)\rangle$, 
and write
$\state{M}{H}\fctrans{\sigma}\state{M_n}{H_n}$. Given $\sigma_1 = 
\langle (\alpha_1,\W_1),\ldots,(\alpha_k,\W_k)\rangle$, $\sigma_2 = 
\langle (\alpha_{k+1},\W_{k+1}),\ldots,(\alpha_n,\W_n)\rangle$,
we write $\sigma_1;\sigma_2$ for $\langle (\alpha_1,\W_1),\ldots,(\alpha_n,\W_n)\rangle$. 
We may also use the notation $\sigma_1;\sigma_2$ when 
$\sigma_1$ or $\sigma_2$ is a single transition.
A central concept in what follows is causal equivalence on traces, a notion that employs
the concept of concurrent transitions:
\begin{definition}{\rm
		Consider an MRPN $(P,T,\A,\A_V,\B,F)$, a reachable state 
		$\state{M}{H}$ and actions $(\alpha_1,\W_1)$ and 
		$(\alpha_2,\W_2)$. Then  $(\alpha_1,\W_1)$ and 
		$(\alpha_2,\W_2)$ are said to be \emph{concurrent} in state 
		$\state{M}{H}$, if for all $u,v\in \A_V$, if $\W_1(u)=A_i$ and 
		$\W_2(v)=B_j$, $A_i,B_j\in M(x)$ then
		$\connected(A_i,M(x))\neq \connected(B_j,M(x))$.
}\end{definition}
Thus, two actions are concurrent when they employ different token instances. This notion
captures when two actions are independent, i.e. the execution
of the one does not preclude the other. Indeed, we may prove the following results.
 \begin{proposition}[Square Property]\label{SP}{\rm
Consider an MRPN $(P,T,\A,\A_V,\B,F)$, a reachable state 
		$\state{M}{H}$ and concurrent actions $(\alpha_1,\W_1)$ and 
		$(\alpha_2,\W_2)$ in $\state{M}{H}$, such that $\state{M}{H}\fctrans{(\alpha_1,\W_1)}\state{M_1}{H_1}$ and $\state{M}{H}\fctrans{(\alpha_2,\W_2)}\state{M_2}{H_2}$.
		Then $\state{M_1}{H_1}\fctrans{(\alpha_2,\W_2)}\state{M'}{H'}$ and $\state{M_2}{H_2}\fctrans{(\alpha_1,\W_1)}\state{M''}{H''}$,
		where $\state{M'}{H'} \asymp \state{M''}{H''}$.}
\end{proposition}
\emph{Proof:} 
It is easy to see that since the two transitions
involve distinct tokens then they can be executed in any order. If,
additionally, $\alpha_1\neq\alpha_2$ or $\alpha_1=\alpha_2$ and they are both
reverse transitions, then the effects imposed on the histories and
the tokens of the transitions  will be independent
and the same in both cases, i.e. $\state{M'}{H'}=\state{M'}{H'}$.
If instead $\alpha_1=\alpha_2$ and $\alpha_1$, 
$\alpha_2$ are not both reverse transitions,
then it is possible that distinct tokens will be assigned to the forward
transition(s). Nonetheless, the sequence of actions executed by each
token instance will be the same in both interleavings and, thus, the
resulting states are causally equivalent.
\proofend

\begin{proposition}[Reverse Transitions are Independent]\label{RTI}{\rm
Consider an MRPN $(P,T,\A,\A_V,\B,F)$, a state 
		$\state{M}{H}$ and enabled reverse actions $(\underline{t_1},\R_1)$ and 
		$(\underline{t_2},\R_2)$  where
		$(\underline{t_1},\R_1)\neq 
		(\underline{t_2},\R_2)$. Then,$(\underline{t_1},\R_1)$ and 
		$(\underline{t_2},\R_2)$ are concurrent.}
\end{proposition}
\emph{Proof:} 
It is straightforward to see that two distinct reverse transitions
employ different tokens. This is because a token instance may
only reverse the last transition occurrence in its history.
Therefore $(\underline{t_1},\R_1)$ and 
		$(\underline{t_2},\R_2)$ satisfy the requirement for
		being concurrent. 
\proofend

We also define two transitions
to be opposite in a certain state as follows:
\begin{definition}{\rm
	Consider an MRPN $(P,T\A,\A_V,\B,F)$ and actions $(\alpha_1,\W_1)$ and 
	$(\alpha_2,\W_2)$. Then  $(\alpha_1,\W_1)$ 
	and $(\alpha_2,\W_2)$ are said to be \emph{opposite} if 
	$\underline{\alpha_1} = \alpha_2$ and, if $\alpha_i = t$ for some $t$,
	for all $a\in\guard{t}$, $\init{\W_i(a)} = \W_{3-i}(a)$.
}\end{definition}

Note that this may arise exactly when the two actions are forward
and reverse executions of the same transition and using the same token
instances. We are now ready to define when two traces are causally
equivalent. 
\begin{definition}\label{co-executions}{\rm  
Consider a reachable state $\state{M}{H}$.
Then
\emph{causal equivalence on traces with respect to} $\state{M}{H}$, denoted by $\sigma_1\asymp_{\state{M}{H}}\sigma_2$,
is the least equivalence relation on traces such that (i)
$\sigma_1= \sigma;(\alpha_1,\W_1);$ $(\alpha_2,\W_2);\sigma'$
where $\state{M}{H}\fctrans{\sigma}\state{M'}{H'}$
and
if  $(\alpha_1,\W_1)$ and $(\alpha_2,\W_2)$ are concurrent in
$\state{M'}{H'}$ then $\sigma_2 = \sigma;(\alpha_2,\W_2);(\alpha_1,\W_1);\sigma'$,
and (ii) if  $(\alpha_1,\W_1)$ and $(\alpha_2,\W_2)$ are opposite transitions
then $\sigma_2 = \sigma;\epsilon;\sigma'$.
}\end{definition}

We may now establish the Parabolic Lemma, which
states that causal equivalence allows the permutation of reverse and forward transitions that have no causal relations between them. 
Therefore, computations are allowed to reach for the maximum freedom of choice going backward and then continue forward.

\begin{lemma}[Parabolic Lemma]\label{PL}{\rm \
Consider an MRPN $(P,T,\A,\A_V,\B,F)$, a reachable state 
		$\state{M}{H}$, and an execution $\state{M}{H} \frtrans{\sigma}\state{M'}{H'}$. Then there exist traces $r,r'$ both forward such that $\sigma\asymp_{\state{M}{H}}\underline{r};r'$ and
		$\state{M}{H} \frtrans{\underline{r};r'}\state{M''}{H''}$ 
		where $\state{M'}{H'}\asymp \state{M''}{H''}$.
}\end{lemma}
\emph{Proof} Following~\cite{DBLP:conf/fossacs/LanesePU20}, given the satisfaction of the Square Property
(Proposition~\ref{SP}) and the independence of reverse transitions 
(Proposition~\ref{RTI}), we conclude that the lemma holds. A proof 
from first principles may also be found in~\cite{KP-2020}.
\proofend

We conclude with Theorem~\ref{multimain} 
stating that two computations beginning in the 
same state lead to equivalent states if and only if the two
computations are causally equivalent. 
This guarantees the consistency of the approach since reversing transitions in causal order is in a sense equivalent
to not executing the transitions in the first place. Reversal does not give rise
to previously unreachable states, on the contrary, it gives rise to causally-equivalent 
states due to different keys being possibly assigned 
to concurrent transitions.
\begin{theorem}\label{multimain}{\rm 
Consider an MRPN $(P,T,\A,\A_V,\B,F)$, a reachable state 
		$\state{M}{H}$, and traces $\sigma_1$,
		$\sigma_2$ such that 
		$\state{M}{H} 
\fctrans{\sigma_1} \state{M_1}{H_1}$ and $\state{M}{H}\fctrans{\sigma_2} 
\state{M_2}{H_2}$. Then, $\sigma_1\asymp_{\state{M}{H}}\sigma_2$ 
if and only if   $\state{M_1}{H_1}\asymp\state{M_2}{H_2}$.
	}
\end{theorem}
\emph{Proof:} Following~\cite{DBLP:conf/fossacs/LanesePU20}, given the satisfaction of the Parabolic Lemma and the fact that the model does not allow
infinite reverse computations, we conclude that the theorem holds. A proof 
from first principles may also be found in~\cite{KP-2020}.
\proofend

\section{Multi Tokens versus Single Tokens}

We now proceed to define Single Reversing Petri Nets  as
MRPNs where each token type corresponds to exactly one token instance. 

\begin{definition}{\rm
	A \emph{Single Reversing Petri Net} (SRPN)  $(P,T,\A,\A_V,\B,F)$ is an 
	MRPN where for all $A\in \A$, $|A|= 1$.

}\end{definition}

Forward and causal-order reversal for SRPNs is defined
as for MRPNs. Consequently, SPRNs are special instances of MRPNs. 
In the sequel, we will show that
for each MRPN there is an ``equivalent'' SRPN. 
To achieve this, similarly to \cite{IndividualTokens}, we will employ
Labelled Transition Systems defined as follows:
\begin{definition}\label{LTS}{\rm  A labelled transition system (LTS) is a tuple
$(Q, E, \rightarrow, I)$ where:
		\begin{itemize}
			\item $Q$ is a countable set of states,
			\item $E$ is a countable set of actions,
			\item $\rightarrow \subseteq Q \times E \times Q$ is 
			the step transition relation, where we write 
			$p \trans{u} q$ for $(p, u, q) \in \rightarrow$, and
			\item $I \in Q$ is the initial state.
		\end{itemize}
}\end{definition}

For the purposes of our comparison, we will employ LTSs in the context
of isomorphism of reachable parts: 

\begin{definition}\label{iso}{\rm 
		Two LTSs $L_1=(Q_1,E_1,\rightarrow_1, I_1)$ and $L_2=(Q_2,E_2,\rightarrow_2, I_2)$
		are isomorphic, written $L_1\cong L_2$, if they
		differ only in the names of their states and events, i.e. if there are bijections
		$\gamma :Q_1 \rightarrow Q_2$ and $\eta :E_1 \rightarrow E_2$ such that 
		$\gamma (I_1)=I_2$, and, for $p,q\in Q_1$, 
		$u \in E_1: \gamma (p)\trans{\eta (u)}_2\gamma (q)$ iff $p\trans{u}_1 q$.
}\end{definition}

The set ${\cal R}(Q)$ of reachable states in $L = (Q, E, \rightarrow, I)$
is the smallest set such that $I$ is reachable and whenever $p$ is 
reachable and $p \trans{u} q$ then $q$ is reachable.
The reachable part of $L$ is the LTS ${\cal{R}} (L) = (R(Q), E, \rightarrow_\R, I)$, where $\rightarrow_\R$ is the part of
the transition relation restricted to reachable states.
We write $L_1 \cong_{\cal{R}} L_2$ if ${\cal{R}}(L_1)$ and ${\cal{R}}(L_2)$ are isomorphic.
To check $L_1 \cong_{\cal{R}} L_2$ it suffices to restrict to subsets of $Q_1$ and 
$Q_2$ that contain all reachable states, and construct an isomorphism 
between the resulting LTSs.

We proceed to give a translation from MRPNs to SPRNs.  First,
we present how an LTS can be associated with an MRPN/SRPN structure.

\begin{definition}{\rm  Let $N = (P,T,\A,\A_V,\B,F)$ be an MRPN (or SRPN)
with initial marking $M_0$.
Then 
${\cal{H}}(N,M_0) = ((P\rightarrow 2^{\A_I \cup \B_I})\times (T\rightarrow
2^\mathbb{N}),
Act, 
\fctrans{}, \state{M_0}{H_0})$ is the LTS associated with $N$.
		\label{LTSs}
}\end{definition}
We may now establish that for any MRPN there exists an SPRN 
with an isomorphic LTS.

\begin{theorem}\label{express}{\rm For every MRPN $N = (P,T,\A,\A_V,\B,F)$ with initial 
marking $M_0$ there exists an SRPN $N' = (P,T',\A',\A_V', \B', F')$  with 
initial marking $M_0'$ such that ${\cal{H}}(N,M_0) \cong_{\cal{R}}
{\cal{H}}(N',M_0')$.
}\end{theorem}
\emph{Proof:} Let $N = (P,T,\A,\A_V,\B,F)$ be an MRPN with initial 
state $\state{M_0}{H_0}$.  We introduce the notation $\W\da$ where for any type-respecting
assignment $\W$, $\W\!\da\!\!(a) = \W(a)\!\da$, that is $\W$ assigns to a variable
in the range of $\W$ the token instance associated to it by $\W$ but
with its history removed. Furthermore, if $f=\W\!\da$ we write $f_s(a)=a_i$ if
$f(a) = (A,i)$.
We construct  $N' = (P,T',\A',\A_V', \B', F')$ with initial
state $\state{M_0'}{H_0'}$ as follows:
\begin{eqnarray*}
	\A'&=&\{A_i \mid \exists (A,i,[])\in M_0(x) \mbox{ for some } x \in P\}\\
	\A_V'&=& \{a_i\mid A_i\in	\A'\}\\
	\B'&=&\{(A_i,B_j)\mid A_i,B_j\in\A', (A,B)\in\B\}\\
	T'&=& \{t_{\W\da}\mid t\in T, \W:\guard{t}\rightarrow \A_I  \mbox{ is a type-respecting assignment } \}\\
	F'(x_1,x_2) &=&
	\{f_s(a) \mid  \exists i\in \{1,2\}, x_i=t_f\in T', \mbox{ and } a\in\guard{t}\}\\
	&\cup& \{(f_s(a),f_s(b))\mid \exists i\in \{1,2\}, x_i=t_f\in T', (a,b)\in \guard{t}\}\\
	M_0'(x) &= &\{
	(A_i,1,[])\mid (A,i,[])\in M_0(x)\}, \;\forall x\in P\\
	H_0'(t) &= &\es, \; \forall t\in T'
	\end{eqnarray*}
The above construction, projects each type $A$ in $N$
to a set of types $A_i$ in $N'$ such that, $A_i\in \A'$ for each instance $(A,i,[])$
of type $A$
in $M_0$. Type $A_i$ contains exactly one element,
initially named $(A_i,1,[])$. Furthermore, for each transition $t\in T$,
we create a set of transitions of the form $t_f\in T'$, to associate
all possible ways in which token/bond instances may be taken as input
by $t$ with a distinct transition that takes as input the combination
of types projected to by the instances.

We now proceed to define bijections $\gamma$ and $\eta$ 
for establishing the homomorphism between the two LTSs.
To simplify the proof, we assume that during the execution
of transitions the enabling assignment is recorded both in the
transition histories, i.e. given a transition $t$ we have $(k,\W)\in H(t)$
signifying that the $k^{th}$ occurrence of $t$ was executed with enabling
assignment $\W$ and also in a token instance $(A,i,xs)$ elements
of $xs$ have the form $(k,t,v,\W\!\da)$ again recording the assignment
that enabled the specific execution of the transition occurrence.
In this setting, it is easy to associate each token instance of $N$
to a token instance of $N'$ as follows, where we write $st(A_i)$ for
the equivalent token instance of $A_i$ in the SPRN $N'$:

$\;\;\;st((A,i,xs))=(A_i,1,ys)$ 

\noindent where, if $xs=[(k^i,t^i,v^i,f^i)]_{1\leq i \leq n}$ then
$ys = [(|\{(k,t^i,v,f)\mid \exists k, v, f \mbox{ s.t.} (k,t^i,v,f)\in ys\}|,t_f^i,f_s(a))]_{1\leq i \leq n}$.

For any reachable state $\state{M}{H}$ in LTS ${\cal{H}}(N,M_0)$, 
we define $\gamma(\state{M}{H})=\state{M'}{H'}$ such that
for all $x\in P$ and $t_f\in T'$
\begin{eqnarray*}
M'(x)&=&\{st(A_i)\mid A_i\in M(x)\}\cup\{(st(A_i),st(B_j))\mid (A_i,B_j)\in M(x)\}\\
H'(t_f) &=& \{1, \ldots, k \mid k=|\{ (i,\R)\in H(t)\mid\R\!\da=f\}|\}
\end{eqnarray*}

Furthermore, given an action $(t,\W)$, we write
\[\eta((t,\W))=(t_{\W\da},\W')\]
where if $\W(a)=A_i$ then $\W'(a_i)=st(A_i)$.

Based on these, we may confirm that there exists an isomorphism between the LTSs ${\cal{H}}(N,M_0)$ and ${\cal{H}}(N',M_0')$ as follows. Suppose $\state{M_m}{H_m}$ is a reachable state of ${\cal{H}}(N,M_0)$ 
	with $\gamma(\state{M_m}{H_m})= \state{M_s}{H_s}$. Two cases exist:
\begin{itemize}
	\item Suppose 
	$\state{M_m}{H_m}\trans{(t,\S)}\state{M_m'}{H_m'}$. This implies that $t$ is a
	forward-enabled transition with forward-enabling assignment $\S$. 
	Consider $\eta(t,\S)=(t_{\S\da},\S')$, as defined above.
	It is easy to
	see that $t_{\S\da}$ is also a forward-enabled transition in $\state{M_s}{H_s}$ with
	forward-enabling assignment $\S'$. Furthermore, if $\state{M_s}{H_s}\trans{(t_{\S\da},\S')}\state{M_s'}{H_s'}$, then 
	\begin{eqnarray*}
		M_s'(x) & = & 
		 (M_s(x)-\bigcup_{a\in F'(x,t_{\S\da})}\connected(\S'(a),M'(x)))\\
	   &\cup& \;\bigcup_{a\in F'(t_{\S\da},x)}\connected(\S'(a),\after(t_{\S\da},\S',M_s)) \oplus(\S',t_\S,k)\\
	   & = & 
		 (M_s(x)-\bigcup_{a\in F(x,t)}\{st(A_i), (st(A_i),st(B_j))\mid A_i, (A_i,B_j)\in\connected(\S(a),M_m(x)))\\
	   &\cup& \;\bigcup_{a\in F(t,x)}\{st(A_i),  (st(A_i),st(B_j))\mid A_i, (A_i,B_j)\in\\
	   & & \hspace{1.5in}\connected(\S(a),\after(t,\S,M_m)) \oplus(\S,t,k)\}
	   \end{eqnarray*}
	   
		where $k=max(\{0\} \cup \{k'|k'\in H(t)\})+1$ and
		\[
		\hspace{1.5 cm}
		H'(t) = \left\{
		\begin{array}{ll}
		H(t) \cup  \{k\}, \hspace{1.2cm} \textrm{ if } t = t_{\S\da}  \; \\
		H(t'), \hspace{2cm} \textrm{ otherwise }
		\end{array}
		\right.			\]
		
	We may see that $\gamma(\state{M_m}{H_m})=\state{M_s}{H_s}$, and the
	result follows. Reversing the arguments, we may also prove the opposite
	direction.
	\item Suppose $\state{M_m}{H_m}\rtrans{(t,\R)}\state{M_m'}{H_m'}$.
	This implies that $t$ is a reverse enabled transition with enabling
	assignment $\R$. 	Consider $\eta(t,\R)=(t_{\R\da},\R')$, as defined above.
	It is easy to
	see that $t_{\R\da}$ is also a reverse-enabled transition in $\state{M_s}{H_s}$ with
	reverse-enabling assignment $\R'$. Furthermore, if 
	$\state{M_s}{H_s}\rtrans{(t_{\R\da},\R')}\state{M_s'}{H_s'}$, then 
	using similar arguments as in the previous case we may confirm that $\gamma(\state{M_m}{H_m})=\state{M_s}{H_s}$. The same holds
	for the opposite direction. This completes the proof.
 \proofend
 \end{itemize}

\begin{figure}[t]
	\centering
	\subfigure[MRPN $N$]{\includegraphics[height=4.05cm]{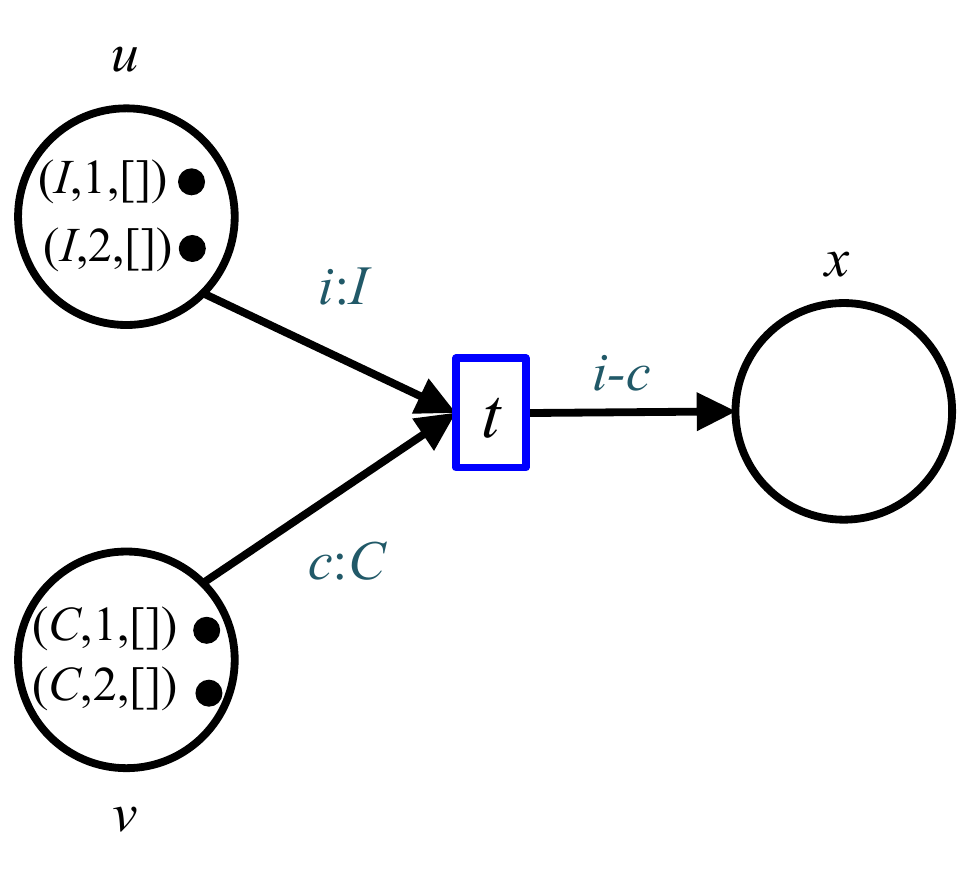}}\hspace{0.4in}
	\subfigure[Equivalent SRPN $N'$]{\includegraphics[height=4.45cm]{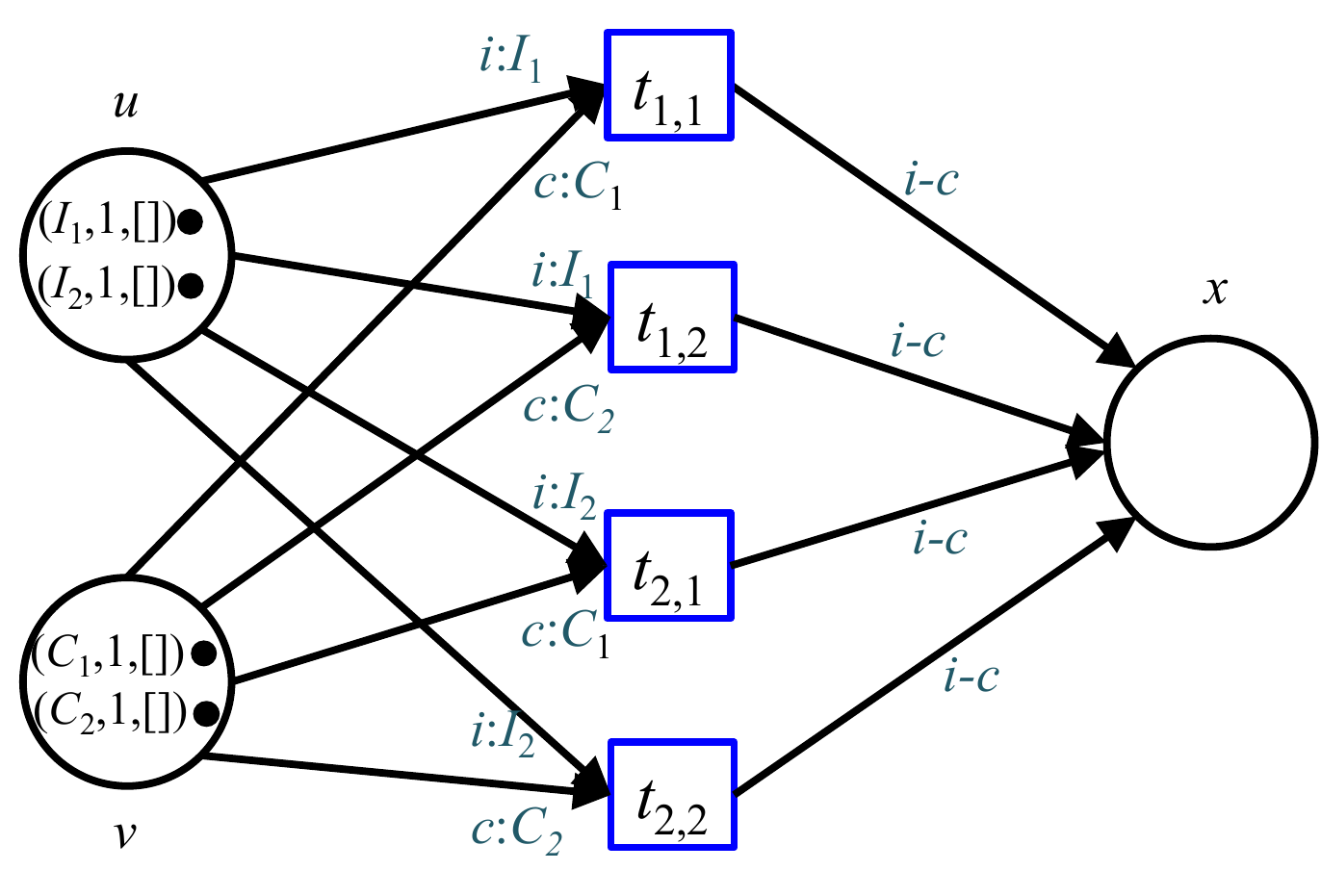}} 
	\caption{Translating MRPNs to SRPNs}
	\label{multiFig}
\end{figure}

In Fig.~\ref{multiFig} we present an MRPN $N$ and its respective SRPN $N'$. From $N$ we
obtain $N'$ by constructing the new token types $I_1,I_2,C_1,C_2$ and
exactly one token instance of each of these types.  The places are the same in both RPN models.
The transitions required for the SRPN are dependent on the types of the variables
required for each MRPN transition and the  token instances representing that type.  
Specifically for each token-instance combination that may fire a transition
in the MRPN, a
respective transition is required in the SRPN. In the example, two token instances of 
type $I$ can be instantiated to variable $i$ and two token instances of type $C$ can
be instantiated to variable $v$. This yields four combinations of token instances
resulting in  four different transitions.

\begin{figure}
	\centering
	\subfigure[LTS of net of Fig. 6(a) ]{\includegraphics[height=4.45cm]{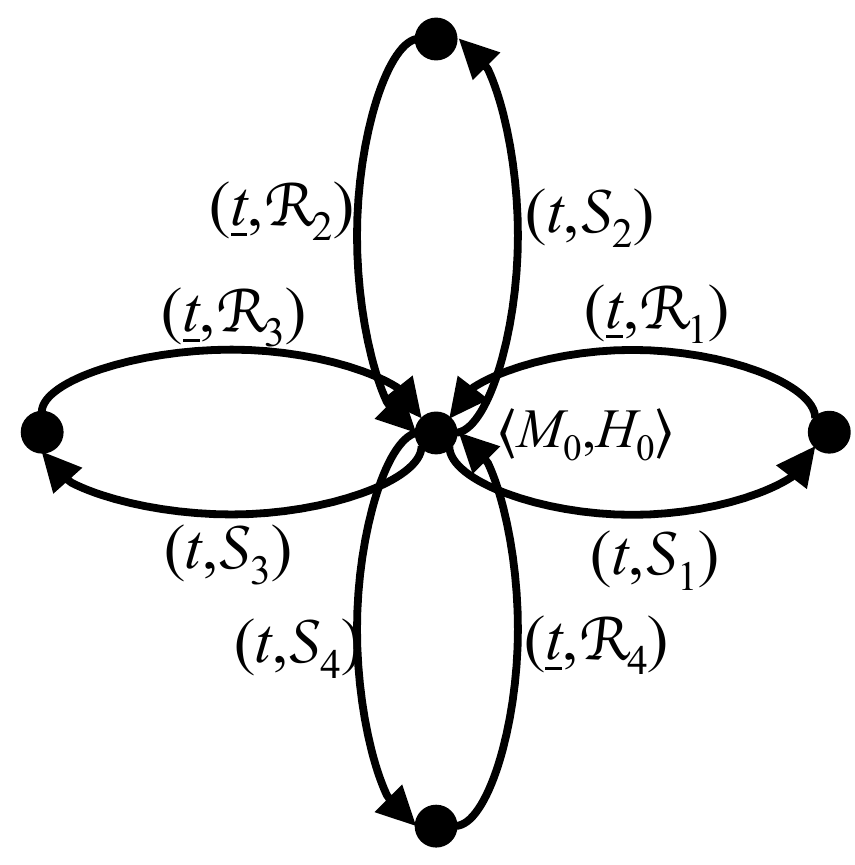}}\hspace{0.6in}
	\subfigure[LTS of net of Fig. 6(b)]{\includegraphics[height=4.45cm]{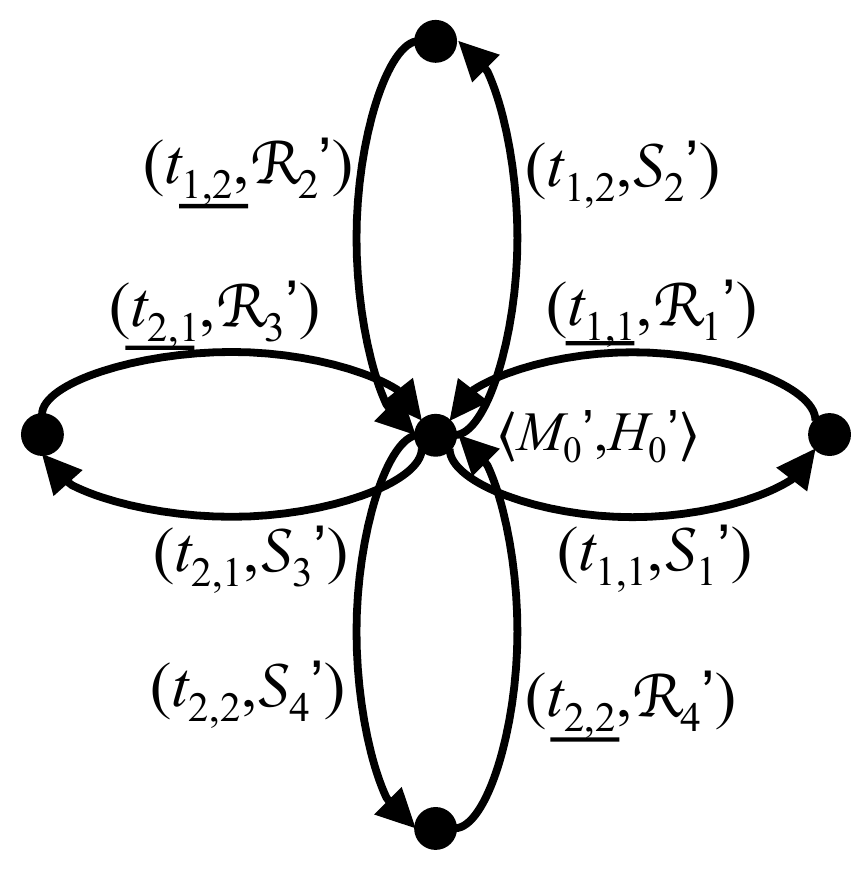}} 
	\caption{Isomorphic LTSs of an MRPN and its SPRN translation}
	\label{EQ-ltss}
\end{figure}

In 
Fig.~\ref{EQ-ltss} we may see the isomorphic LTSs of the two RPNs, where
\begin{eqnarray*}
t_{1,1}=t_{\S_1\da}&& \underline{t_{1,1}}=\underline{t_{\R_1\da}}\\
t_{1,2}=t_{\S_2\da}&& \underline{t_{1,2}}=\underline{t_{\R_2\da}}\\
t_{2,1}=t_{\S_3\da}&& \underline{t_{2,1}}=\underline{t_{\R_3\da}}\\
t_{2,2}=t_{\S_4\da}&& \underline{t_{2,3}}=\underline{t_{\R_4\da}}
\end{eqnarray*}
and the enabling assignments of the actions in the two LTSs are 
\begin{eqnarray*}
\S_1(i)=(I,1,[]),&& \S_1(c) = (C,1,[])\\
\R_1(i)=(I,1,[(t,1,i)]), &&\R_1(c) = (C,1,[(t,1,c)])\\
\S_2(i)=(I,1,[]), &&\S_2(c) = (C,2,[])\\
\R_2(i)=(I,1,[(t,1,i)]), &&\R_2(c) = (C,2,[(t,1,c)])\\
\S_3(i)=(I,2,[]), &&\S_3(c) = (C,1,[])\\
\R_3(i)=(I,2,[(t,1,i)]), &&\R_3(c) = (C,1,[(t,1,c)])\\
\S_4(i)=(I,2,[]), && \S_4(c) = (C,2,[])\\
\R_4(i)=(I,2,[(t,1,i)]), &&\R_4(c) = (C,2,[(t,1,c)])
\end{eqnarray*}
and
\begin{eqnarray*}
\S_1'(i)=(I_1,1,[]),&& \S_1(c) = (C_1,1,[])\\
\R_1(i)=(I_1,1,[(t,1,i)]), &&\R_1(c) = (C_1,1,[(t,1,c)])\\
\S_2(i)=(I_1,1,[]), &&\S_2(c) = (C_2,1,[])\\
\R_2(i)=(I_1,1,[(t,1,i)]), &&\R_2(c) = (C_2,1,[(t,1,c)])\\
\S_3(i)=(I_2,1,[]), &&\S_3(c) = (C_1,1,[])\\
\R_3(i)=(I_2,1,[(t,1,i)]), &&\R_3(c) = (C_1,1,[(t,1,c)])\\
\S_4(i)=(I_2,1,[]), && \S_4(c) = (C_2,1,[])\\
\R_4(i)=(I_2,1,[(t,1,i)]), &&\R_4(c) = (C_2,1,[(t,1,c)])
\end{eqnarray*}

\section{Conclusions}
This paper presents an extension of RPNs with
multiple tokens of the same type based on the
individual token interpretation. The individuality of tokens is enabled by 
recording their causal path, while the semantics allows
identical tokens to fire any eligible transition when going forward, but
only the transitions they have been previously involved in when
going backward. We have presented a semantics for causal-order reversibility,
which unlike the semantics presented in~\cite{RPNscycles} is purely
local and requires no global control.
Another contribution of the paper is a result 
illustrating that introducing multiple tokens in the model does not increase 
its expressive power. Indeed, for every MRPN we may construct an equivalent 
SRPN, which preserves its computation. In related work~\cite{KP-2020},
MRPNs have also been associated with backtracking and out-of-causal-order 
semantics and it was shown that in all settings MRPNs are equivalent to the 
original RPN model.

In our current work 
we are developing
a tool for simulating and verifying RPN models~\cite{ASPtoRPNs},
which we aim to apply towards the analysis
of resource-aware systems. Our 
experience in applying RPNs in the context of wireless
communications~\cite{RC19} has illustrated that resource management 
can be studied and understood in terms of RPNs since, 
along with their visual
nature, they offer a number of features, such as 
token persistence, that is especially relevant in 
these contexts. In future work, we would like
to further apply our framework in the specific fields
as well as in the field of long-running transactions.

\bibliographystyle{eptcs}
\bibliography{References}
\end{document}